\documentclass[authoryear,preprint,12pt]{elsarticle}
\usepackage{amssymb}

\usepackage{amsmath}

\usepackage{amsthm}
\usepackage{lineno}
\usepackage[english]{babel}
\usepackage[utf8]{inputenc}
\usepackage{xcolor}
\usepackage[ruled,vlined]{algorithm2e}
\usepackage{float}
\usepackage{multicol}
\usepackage{booktabs}
\usepackage{graphicx}
\usepackage{colortbl}
\usepackage[hidelinks]{hyperref}
\usepackage{adjustbox}
\usepackage{geometry}
\usepackage{natbib}
\usepackage{cleveref}
\usepackage{comment}
\usepackage{rotating}
\usepackage{tikz}
\usepackage{tikz-qtree}
\usetikzlibrary{arrows.meta}
\usetikzlibrary{shapes.geometric}
\usepackage{tikzscale}
\usepackage{arydshln}
\usepackage{tabularx}
\usepackage{mathtools}
\usepackage{subcaption}
\usepackage[normalem]{ulem}
\usepackage{mathrsfs}
\usepackage{longtable}
\usepackage{rotating}

\newtheorem{prop}{Proposition}
\newtheorem{lemma}{Lemma}
\newdefinition{remark}{Remark}
\newdefinition{assumption}{Assumption}
\newdefinition{proc}{Procedure}
\newdefinition{defi}{Definition}

\newcommand{\V}{\Tilde{\mathbf{V}}^{-1}}
\newcommand{\TV}{\Tilde{\mathbf{V}}}
\newcommand{\Xlong}[1]{\mathbf{X}_{#1}(\mathbf{X}_{#1}^{\top} \mathbf{X}_{#1})^{-1}\mathbf{X}_{#1}^{\top}}
\newcommand{\sqrho}{\frac{\rho^2}{\rho^2-1}}
\newcommand{\sirho}{\frac{\rho}{\rho^2-1}}
\newcommand{\divsig}{\frac{1}{\sigma^2}}
\newcommand{\divsigtwice}{\frac{1}{2\sigma^2}}

\newcommand{\X}{\mathbf{X}}

\newcommand{\M}{\mathcal{M}}
\newcommand{\js}{\hat{\ell}}
\newcommand{\C}[1]{\mathcal{C}_{T|#1}(\theta_{#1},\Tilde{\mathbf{y}})}
\newcommand{\CBOLD}[1]{\mathcal{C}_{T|#1}(\beta_{#1,ijkc_k},\boldsymbol{y}_{obs,ij})}
\newcommand{\bbeta}{\boldsymbol{\zeta}}
\newcommand{\G}{\mathbf{G}}
\newcommand{\arho}{1-\sqrho}
\newcommand{\brho}{\sirho}
\newcommand{\crho}{1-2\sqrho}
\newcommand{\Y}{\Tilde{\mathbf{y}}}
\newcommand{\TW}{\Tilde{\mathbf{w}}}
\newcommand{\BW}{\Bar{\mathbf{w}}}

\newcommand{\Sij}{\mathbf{S}_{ij}}
\newcommand{\HS}{\hat{\mathbf{S}}_{ij}}

\journal{Computational Statistics \& Data Analysis}

\begin{document}

\begin{frontmatter}


\title{Identifying rapid changes in the hemodynamic response in event-related functional magnetic resonance imaging} 

\author[label1]{Friederike Preusse}\corref{cor1}
\ead{preusse@uni-bremen.de}
\author[label1]{Thorsten Dickhaus}
\author[label2]{André Brechmann}

\cortext[cor1]{Corresponding author}
\affiliation[label1]{organization={University of Bremen, Institute for Statistics},
            addressline={Bibliothekstrasse 5}, 
            city={Bremen},
            postcode={28359}, 
            country={Germany}}
\affiliation[label2]{organization={Leibniz-Institute for Neurobiology},
            addressline={Brenneckestrasse 6}, 
            city={Magdeburg},
            postcode={39118}, 
            country={Germany}}

\begin{abstract}

The hemodynamic response (HR) in event-related functional magnetic resonance imaging is typically assumed to be stationary. While there are some approaches in the literature to model nonstationary HRs, few focus on rapid changes. In this work, we propose two procedures to investigate rapid changes in the HR. Both procedures make inference on the existence of rapid changes for multi-subject data. 
We allow the change point locations to vary between subjects, conditions and brain regions. 
The first procedure utilizes available information about the change point locations to compare multiple shape parameters of the HR over time. In the second procedure, the change point locations are determined for each subject separately. To account for the estimation of the change point locations, we propose the notion of post selection variance. The power of the proposed procedures is assessed in simulation studies. We apply the procedure for pre-specified change point locations to data from a category learning experiment.

\end{abstract}


\end{frontmatter}

%\linenumbers
\section{Introduction}
In the analysis of event-related functional magnetic resonance imaging (fMRI) data it is common to assume that the brain's reaction to a given type of stimulus is time invariant. In other words, the hemodynamic response (HR) to a type of events, called a condition, is considered to be stationary. 
However, in several settings this stationarity assumption can be violated. For example, nonstationarity of the HR may arise due to changes in the emotional state, changes in attention or because of learning \citep{Muheialdin2014, Poldrack2001, Ischebeck2006}. 

Different approaches to account for nonstationarity in the HR have been proposed.
\citet{Buchel1998} suggest to include available information about the changes when modeling the assumed HR, such as response time \citep{Grinband2008, Mumford2023} or task performance \citep{Morita2019}. This approach requires prior knowledge about the influence these variables have on the shape of the HR.
If no prior information about the changes in the HR is available, the variation of the HR magnitude can be modeled using weights \citep{Donnet2006, Hinrich2000} or by treating each stimulus as a separate condition \citep{Napadow2009}. \citet{Kalus2015} and \citet{Park2020} utilize penalized spline regression to allow the shape of the HR to vary smoothly over time. 
Besides changing smoothly, the HR can also vary rapidly, for example, when emotions change \citep{Robinson2010, Candemir2023}. Rapid changes have also been found in resting-state fMRI \citep[see e.g.,][]{Aston2012, Ghannam2024} and in functional connectivity analyses \citep[see e.g.,][]{Monti2018, Zhao2022}. 
For task-related fMRI experiments, \citet{Lindquist2007} and \citet{Robinson2010} utilize change point analysis procedures to study state-related changes in the baseline of the fMRI signal. \citet{Candemir2023} utilize neural networks to make inference on the existence of rapid changes in the overall shape of the HR for event-related fMRI. 
Alternatively, the observed fMRI signal, called the BOLD (blood-oxygenation-level dependent) signal, can be split into pre-defined time segments \citep{Milham2003, Menz2006, Morrot2013}. Within each segment, the HR is modeled under the assumption of stationarity. This allows for the comparison of the overall shape of the HR over time and over multiple subjects.

The existing procedures to investigate changes in the shape of the HR have some limitations. \citet{Kalus2015, Park2020} and \citet{Candemir2023} make inference at the subject level, thus results are subject specific. \citet{ Hinrich2000, Donnet2006,Lindquist2007} and \citet{Robinson2010} investigate changes in only one aspect of the shape of the HR, such as the magnitude of the HR or its baseline. When splitting the BOLD signal, the rapid changes are assumed to occur simultaneously in all brain regions and simultaneously for all conditions. 

In this paper, we aim to overcome these limitations. To this end, we propose two novel analysis pipelines to make inference on the existence of rapid changes in the HRs to multiple conditions during an event-related fMRI experiment.
In contrast to existing methods, the change point locations may vary between conditions, brain regions, and subjects. Furthermore, we make inference on rapid changes in the HR at the group level rather than at the subject level. 

We differentiate between two scenarios. In the first scenario, prior information about the change point locations is available. 
In this case, we propose a procedure that allows for an in-depth investigation of the changes in the HR. 
That is, for each condition and brain region, we investigate rapid changes in several shape parameters that describe the HR. 
The second proposed procedure does not require prior information about the change point locations. 
We determine the change point locations for each subject individually, while making inference on the existence of rapid changes at the group level. We define and utilize the post selection variance to account for the estimation of the change point locations. We define the post selection variance in a general manner, and its application is not limited to neuroimaging data.

The remainder of this work is structured as follows. In the next section, we introduce notation using the example of stationary HRs. 
The proposed methodology to make inference on rapid changes in the HRs is introduced in Sections \ref{sec:methodology_known_cp} and \ref{sec:methodology_unknowncp} for pre-specified and undefined change point locations, respectively. 
We demonstrate the performance of the procedure for pre-specified change point locations by means of a simulation study in Section \ref{sec:simulation}.
The methodology is applied to data from an event-related category learning experiment in Section \ref{sec:application}. We conclude with a discussion of the proposed procedure and an outlook to future work. Proofs and additional results are provided in the appendix.

Parts of the investigation have been presented at the $6$th International Conference for Statistics in Theory and Application (ICSTA'24), see also \url{https://international-aset.com}.

\section{Notation and Preliminaries}
\label{sec:fmri_intro}
Throughout, we denote matrices by upper case bold letters and vectors by lower case bold letters. Unless stated otherwise, letters not in bold refer to a scalar. We refer to an entry of a matrix or vector by indicating the entry's position in square brackets.

In the following we introduce the notation for event-related fMRI analysis for the case of stationary HRs. Exhaustive information regarding the analysis of fMRI data under the stationarity assumption can be found in \citet{ Huettel2004, Lindquist2008, Lindquist2009, poldrack_mumford_nichols_2011} and \citet{Thirion2017}, among others. 

In the case of multi-subject experiments, fMRI data analysis is generally done in a two-step procedure, the subject level analysis and the group level analysis.
At the subject level, the BOLD time series of length $T$ of subject $i$, $i=1,\ldots,n$, observed at voxel $j$, $j=1,\ldots,J$, is denoted by $\boldsymbol{y}_{ij}\in \mathbb{R}^T$. 
Let the matrix $\boldsymbol{Z}_{ik}\in\mathbb{R}^{T\times G}$ account for the rudimentary HR to the $k$-th condition, with $k=1,\ldots,K$. The rudimentary HR is the convolution of a hemodynamic response function (HRF) and the condition $k$ onset time series $\boldsymbol{u}_{ik}$, $\boldsymbol{u}_{ik}\in \{0,1\}^T$. 
Denote by $\boldsymbol{\Tilde{Z}}_{i}$ a matrix containing confounding variables, such as motion parameters or the baseline. Under the assumption of stationarity of the HRs, $\boldsymbol{y}_{ij}$ is modeled as
\begin{equation}
\label{eq:BOLD_at_time_t}
    \boldsymbol{y}_{ij}=\sum_{k=1}^K\boldsymbol{Z}_{ik}\boldsymbol{\beta}_{ijk}+\boldsymbol{\Tilde{Z}}_{i}\boldsymbol{\delta}_{ij}+\boldsymbol{\epsilon}_{ij},
\end{equation}
where $\boldsymbol{\beta}_{ijk}$ and $\boldsymbol{\delta}_{ij}$ are the vectors of regression coefficients corresponding to $\boldsymbol{Z}_{ik}$ and $\boldsymbol{\Tilde{Z}}_{i}$, respectively. 
The noise is modeled by $\boldsymbol{\epsilon}_{ij}\sim \mathcal{N}_T(\boldsymbol{0}_T,\sigma_{\epsilon_{ij}}^2 \boldsymbol{V})$, where $\boldsymbol{0}_T$ denotes the null vector of length $T$. 
The covariance matrix $\boldsymbol{V}$ expresses an auto regressive (AR) process of first or second order.

\begin{figure}[ht]
    \centering
    \includegraphics[width=0.6\textwidth]{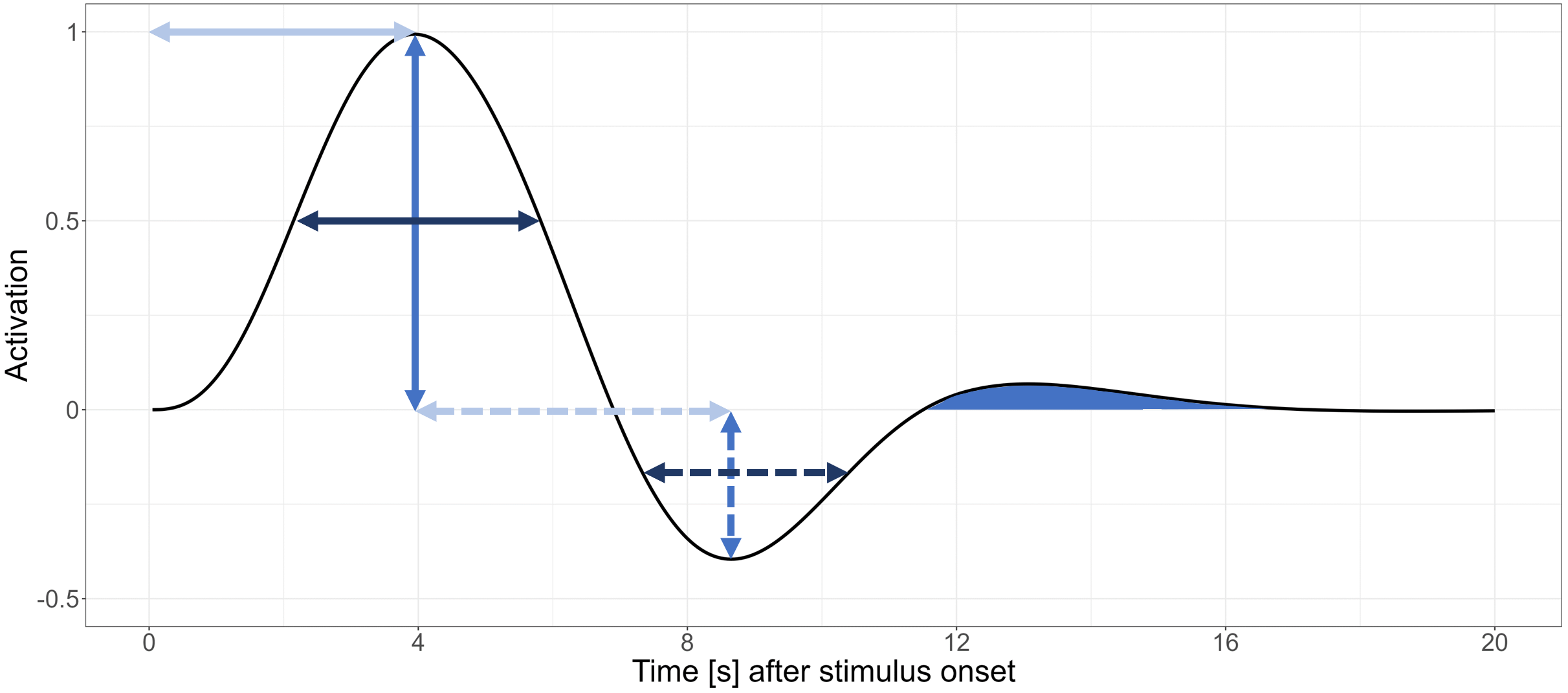}
   \caption{Exemplary shape of the hemodynamic response with stimulus onset at $t=0$ seconds. Different shape parameters of the HR are displayed: the peak magnitude, $PM$, (solid blue arrow), nadir amplitude, $NA$, (dotted blue arrow), time to peak, $TTP$, (solid light blue arrow), time peak to nadir, $TPN$, (dotted light blue arrow) as well as full width at half maximum, $FWHM$, (solid dark blue arrow) and full width at half nadir, $FWHN$, (dotted dark blue arrow). The area under the curve, $AUC$, is shaded in blue.}
   \label{fig:shape_HRF}
\end{figure}

The definition of $\boldsymbol{Z}_{ik}$ depends on the underlying HRF model. We only consider HRF models which are a linear combination of $G$ basis functions, such as the canonical HRF, the flexible linear optimal basis sets (FLOBS) \citep{Woolrich2004} or the finite impulse response model \citep[c.f.][]{LindquistWagner2007}. The HR to condition $k$ is estimated as
\begin{equation*}
    \label{eq:est_HR}
    \widehat{hr}_{ijk}=\boldsymbol{\mathcal{B}}\hat{\boldsymbol{\beta}}_{ijk},
\end{equation*} where $\boldsymbol{\mathcal{B}}\in\mathbb{R}^{T'\times G}$ denotes the matrix containing the $G$ basis functions determined by the HRF model. The length $T'$ of the basis functions is not fixed. However, since the HR to one stimulus is considered to last up to 20 seconds, it is sensible to choose $T'$ such that the estimated HR spans (at least) 20 seconds. 
The shape of $\widehat{hr}_{ijk}$ is described by the shape parameter $\hat{\gamma}_{ijk}$. Different shape parameters are illustrated in Figure \ref{fig:shape_HRF}. The variance of $\hat{\gamma}_{ijk}$ is denoted by $\sigma^2_{\hat{\gamma}_{ijk}}$. If $\sigma^2_{\hat{\gamma}_{ijk}}$ cannot be derived explicitly, it can be numerically approximated. 
We describe such an approach in \ref{sec:est_subject_variance}.

At the group level, the vector of the shape parameters of interest across all subjects is denoted by $\hat{\boldsymbol{\gamma}}_{jk}=(\hat{\gamma}_{1jk},\ldots,\hat{\gamma}_{njk})^\top$.  
This vector is modeled by
\begin{equation}
    \label{eq:group_level_stat}
    \hat{\boldsymbol{\gamma}}_{jk}=\eta_{jk}\boldsymbol{1}_n+\boldsymbol{\xi}_{jk},
\end{equation} where $\eta_{jk}$ denotes the one-dimensional shape parameter at the group level and $\boldsymbol{1}_n$ denotes the vector of length $n$ with all entries equal to 1. The noise at the group level is denoted by $\boldsymbol{\xi}_{jk}\in\mathbb{R}^{n}$, with $\boldsymbol{\xi}_{jk}\sim \mathcal{N}_n(\boldsymbol{0}_n,\boldsymbol{\Sigma}_{jk})$ and $\boldsymbol{\Sigma}_{jk}=\sigma^2_{B_{jk}}\mathbf{I}_n+\boldsymbol{\Sigma}_{W_{jk}}$. Here, $\sigma^2_{B_{jk}}$ denotes the between subject variance, $\mathbf{I}_n$ denotes the $(n\times n)$ identity matrix and $\boldsymbol{\Sigma}_{W_{jk}}=diag(\sigma^2_{\hat{\gamma}_{1jk}},\ldots,\sigma^2_{\hat{\gamma}_{njk}})$ denotes the within subject variance matrix \citep{Mumford2006}. 
The null hypothesis of no difference between two conditions, i.e., $H_{j}:\eta_{jk_1}=\eta_{jk_2}$, is of interest for us at every voxel $j\in\{1,\ldots,J\}$. To account for the resulting multiple testing problem, error measures such as the false discovery rate ($FDR$) or family-wise error rate ($FWER$) are utilized. The random variable corresponding to the number of rejected true null hypotheses, i.e., false discoveries, is denoted by $\mathscr{V}$. Let $\mathscr{R}$ denote the random variable corresponding to the number of all rejected null hypotheses. The $FDR$ is the expected value of the false discovery proportion $FDP=\mathscr{V}/(\mathscr{R}\vee 1)$ \citep{BenjaminiHochberg}.  
The $FWER$ is defined as the probability of at least one false discovery. 
We refer to \citet{dickhaus_2014} for an exhaustive overview of multiple testing procedures. 

Many fMRI studies aim at finding clusters of spatially contiguous voxels in which the activation differs between conditions. Such a cluster is called a region of interest (ROI). To determine ROIs, we can employ simultaneous confidence bounds for the true discovery proportion $TDP=1-FDP$ \citep[see e.g.,][]{Rosenblatt2018, Vesely2023}. This approach allows us to define ROIs which include at least a desired proportion of true discoveries with a pre-defined probability.

\section{Pre-specified change point locations}
\label{sec:methodology_known_cp}
In this section, we propose a procedure to make inference on rapid changes in the shape of the HR if reliable information about the change point locations is available.
For example, such information could be based on behavioral data or the experimental design. 
Instead of considering changes in only one shape parameter of interest, we are interested in several shape parameters.
We develop our methodology for pre-defined ROIs and use the average BOLD signal within these ROIs to increase the signal to noise ratio (SNR) \citep{Huettel2004}.
However, for the procedure itself the size of the ROIs is of no relevance. Therefore, the proposed procedure can be applied at the voxel level as well.

We derive the procedure under the following three assumptions:
\begin{assumption}
    \label{as:confound_stationary}
    The confounding effects are stationary.
\end{assumption}

\begin{assumption}
\label{as:rapid_change}
The HR to a given condition is stationary within time segments and changes rapidly between the time segments.
\end{assumption}

\begin{assumption}
\label{as:number_cp_fixed}
    The number of change points is identical for all subjects and known a priori.
\end{assumption}

We call a time segment in which the HR to a given condition is stationary a stationary segment. Stationary segments are defined by the change point locations. They are considered to be condition specific and may vary between ROIs. 

\subsection{Subject level analysis}
To investigate rapid changes in $Q$ shape parameters per condition, we split the condition onset time series $\boldsymbol{u}_{ik}$, $k=1,\ldots,K$, at the change point locations. This results in stationary, segment-specific condition onset time series. By splitting $\boldsymbol{u}_{ik}$ before convolution with the HRF, we account for possible overlaps of the HRs in consecutive stationary segments. 
Denote by $\boldsymbol{\psi}_{ijk}$ %$=\{\psi_{ijk[1]},\ldots,\psi_{ijk[C_{jk}]}\}$
the set of the $C_{jk}$ change point locations for subject $i$, region $j$ and condition $k$. Note that $\boldsymbol{\psi}_{ijk}=\emptyset$ is allowed. 
We define the stationary, segment-specific condition onset time series as
\begin{equation*}
\label{eq:dummy_onset_time_series}
    \boldsymbol{u}_{ik(\boldsymbol{\psi}_{ijk}[c_{jk}-1]:\boldsymbol{\psi}_{ijk}[c_{jk}]-1)}[t]  = \begin{cases}
        \boldsymbol{u}_{ik}[t] \qquad &\text{if } t\in \{\boldsymbol{\psi}_{ijk}[c_{jk}-1],\boldsymbol{\psi}_{ijk}[c_{jk}]-1\}\\
        0 & \text{else,}
    \end{cases}
\end{equation*} for $c_{jk}=1,\ldots,C_{jk}+1$, with $\boldsymbol{\psi}_{ijk}[0]=1$ and $\boldsymbol{\psi}_{ijk}[C_{jk}+1]=T+1$. In the following, we use $c_{jk}$ to indicate the $c$-th change point for condition $k$ and ROI $j$ as well as the $c$-th stationary segment of condition $k$ in ROI $j$, leaving distinction to context. Furthermore, for ease of notation, we only consider the case $C_{1k}=\ldots=C_{Jk}$ explicitly, and notationally omit the subscript $j$.
We model the BOLD signal $\boldsymbol{y}_{ij}$ as
\begin{align}
    \label{eq:BOLD_nonstat_cp}
    \boldsymbol{y}_{ij} = \sum_{k=1}^{K} \sum_{c_k=1}^{C_{k}+1} \boldsymbol{Z}_{ijkc_k}\boldsymbol{\beta}_{ijkc_k}+\boldsymbol{\Tilde{Z}}_{ij}\boldsymbol{\delta}_{ij}+ \epsilon_{ij},\\
    \boldsymbol{Z}_{ijkc_k}=HRF\otimes \boldsymbol{u}_{ik(\boldsymbol{\psi}_{ijk}[c_k-1]:\boldsymbol{\psi}_{ijk}[c_k]-1)},\nonumber
\end{align}
where $\otimes$ denotes convolution.
Note that the regression coefficients given in Eq.~\eqref{eq:BOLD_nonstat_cp} are not time dependent. Therefore, we can use the procedures described in Section \ref{sec:fmri_intro} for further analysis. The shape of the segment-specific HR, $hr_{ijkc_k}$, is described by the shape parameters $\gamma_{ijkc_k}^{[1]},\ldots,\gamma_{ijkc_k}^{[Q]}$. 
\subsection{Group level analysis}
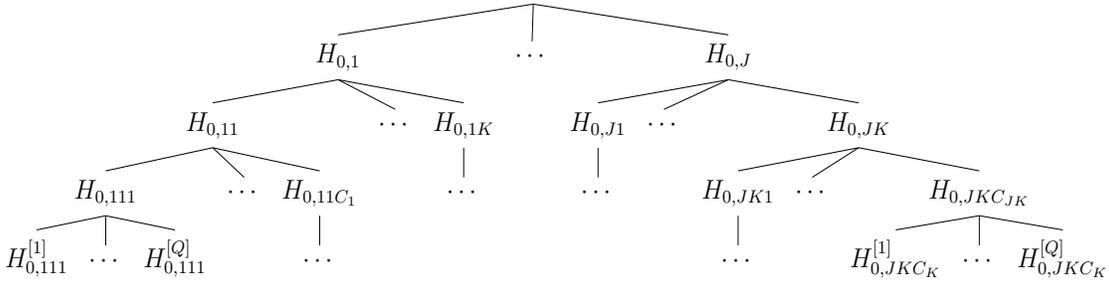
\begin{figure}[ht]
\begin{center}
	\resizebox{15 cm}{4 cm }{\begin{tikzpicture}
    \Tree [.\node (level0-right) {};
        [.$H_{0,1}$
            [.$H_{0,11}$
                [.$H_{0,111}$ $H_{0,111}^{[1]}$ $\cdots$ $H_{0,111}^{[Q]}$ ]
                $\cdots$
                [.$H_{0,11C_{1}}$ $\cdots$
                ]
            ]
            $\cdots$ 
            [.$H_{0,1K}$
                $\cdots$
            ]
        ]
        $\cdots$
        [.$H_{0,J}$ 
            [.$H_{0,J1}$ $\cdots$
            ] 
            $\cdots$
            [.$H_{0,JK}$ 
                [.$H_{0,JK1}$ $\cdots$
                ]
                $\cdots$
                [.$H_{0,JKC_{JK}}$ $H_{0,JKC_{K}}^{[1]}$ $\cdots$ $H_{0,JKC_{K}}^{[Q]}$
                ]
            ]
        ]
    ]
\end{tikzpicture}
%\begin{tikzpicture}
%    \Tree [.\node (level0-right) {Brain};
%        [.$H_{0,1}$
%            [.$H_{0,11}$
%                [.$H_{0,111}$ $H_{0,111}^{[1]}$ $\cdots$ $H_{0,111}^{[Q]}$ ]
%                $\cdots$
%                [.$H_{0,11C_{11}}$ $\cdots$
%                ]
%            ]
%            $\cdots$ 
%            [.$H_{0,1K}$
%                $\cdots$
%            ]
%        ]
%        $\cdots$
%        $H_{0,J}$ 
%    ]
%\end{tikzpicture}
}
	\caption{The hierarchical structure of the hypotheses. The null hypothesis of no change is tested in $J$ regions with $K$ conditions per region, $C_k$ change points per condition and $Q$ shape parameters per change point.
    The hierarchical structure has four levels: the first being the regions, the second being the conditions. The third level includes the stationary segments defined by the change points. The elementary hypotheses corresponding to the $Q$ shape parameters make up the fourth level.}
	\label{figure:Tree_2}
\end{center}
\end{figure}

At the group level, we are interested in the null hypotheses of no change between the stationary segments for each shape parameter, i.e., \begin{equation*}
    H_{jkc_k}^{[q]}:\eta_{jk(c_k+1)}^{[q]}=\eta_{jkc_k}^{[q]},
\end{equation*} for all $q\in\{1,\ldots,Q\}$, $c_k\in\{1,\ldots,C_{k}\}$, $k\in\{1,\ldots,K\}$ and $j\in\{1\ldots,J\}$. 
These hypotheses are called elementary null hypotheses. The elementary null hypotheses can be grouped. Denote by $H_{jkc_k}$ the null hypothesis of no change in the HR to condition $k$ at change point $c_k$ in ROI $j$. The null hypothesis $H_{jkc_k}$ is true if and only if all $H_{ jkc_k}^{[q]}$, $q=1,\ldots,Q$, are true. Hence, $H_{jkc_k}$ is called the parent null hypothesis of $H_{ jkc_k}^{[q]}$, $q=1,\ldots,Q$.  Analogously, the null hypothesis of no change in condition $k$ in ROI $j$, $H_{jk}$, is the parent null hypothesis of $H_{jkc_k}$, $c_k=1,\ldots,C_k$. 
Grouping of the hypotheses continues until all hypotheses are descendants of the node null hypothesis of no change in any region. The full hierarchical structure of the hypotheses is visualized in Figure \ref{figure:Tree_2}. 

We can exploit the hierarchical structure of the hypotheses to reduce the number of elementary hypotheses to be tested. A suitable approach, which controls the so-called selective $FDR$ ($sFDR$), is given by \citet{Bogomolov2020}. At each level of the hierarchy, the $sFDR$ is the weighted average of the $FDR$ in the families under consideration. This procedure requires independence between the families of hypotheses at each level of the hierarchical structure. This requirement is not met, for example, when the proposed procedure is applied at the voxel level, since the HRs in contiguous voxels are positively dependent. A procedure that allows for an arbitrary dependency structure at each level of the hierarchy is proposed by \citet{GoemanFinos2012}. Their so-called inheritance procedure controls the $FWER$ and is potentially less powerful than the approach by \citet{Bogomolov2020} \citep[c.f.][]{BenjaminiHochberg}.

The proposed methodology to include pre-specified change point locations in the analysis of event-related fMRI is summarized in Procedure \ref{proc:known_cp}.
\begin{proc}{Pre-specified change point locations}\\
    \label{proc:known_cp}
\textbf{Input}\\
 For each subject $i=1,\ldots,n$, ROI $j=1,\ldots,J$ and condition $k=1,\ldots,K$:
 \begin{enumerate}
            \item[-] Pre-processed BOLD signal $\boldsymbol{y}_{obs, ij}$
             \item[-] Condition onset time series $\boldsymbol{u}_{ik}$
            \item[-] Change point locations $\boldsymbol{\Psi}_{ij}=\{\boldsymbol{\psi}_{ij1},\ldots,\boldsymbol{\psi}_{ijK}\}$
        \end{enumerate}
 \textbf{Subject level}\\
 For each subject $i$, each ROI $j$, each condition $k$ and each change point $c_k$:
        \begin{enumerate}
            \item Specify $\boldsymbol{u}_{ik[\psi_{ijk}[c_k-1]:(\psi_{ijk}[c_k]-1)]}$.
            \item Estimate $\boldsymbol{\beta}_{ijkc_k}$ given in Eq.~\eqref{eq:BOLD_nonstat_cp}.
            \item Compute $\widehat{hr}_{ijkc_k}$ and the corresponding shape parameters $\hat{\gamma}_{ijkc_k}^{[q]}$, $\forall q\in\{1,\ldots,Q\}$.
            \item Compute $\hat{\sigma}_{\hat{\gamma}_{ijkc_k}^{[q]}}^2$, $q=1,\ldots,Q$.
        \end{enumerate}
\textbf{Group level}\\
For each ROI $j$, each condition $k$ and each shape parameter $q$:
        \begin{enumerate}
            \item For every $c_k=1,\ldots,C_k+1$:\\
            Estimate $\eta_{jkc_k}^{[q]}$ and $\boldsymbol{\Sigma}_{jkc_k}^{[q]}$.
            \item For every $c_k=1,\ldots, C_k$:\\
            Compute the p-value corresponding to the null hypotheses \\$H_{jkc_k}: \eta_{jk(c_k+1)}^{[q]}=\eta_{jkc_k}^{[q]}$.
        \end{enumerate}    
Determine a set of null hypotheses to be rejected.\\
%\vspace{1.5pt}\\
\textbf{Output} \\
 Set of shape parameters that change significantly for every change point per condition and per ROI.
\end{proc}

\begin{remark} 
As mentioned above, the proposed procedure can be applied to the BOLD signal at the voxel level. This allows us to determine regions of contiguous voxels in which the shape of the HR to a given stimulus changes rapidly over time. To this end, we need a map of the brain with one p-value per voxel. For voxel $j$, $j=1,\ldots,J$, this p-value may correspond to $H_{j}$ or one of its descendants.
Then, ROIs can be determined as described in Section \ref{sec:fmri_intro}.
\end{remark}

\section{Undefined change point locations}
\label{sec:methodology_unknowncp}
If no reliable information about the change point locations $\boldsymbol{\Psi}_{ij}$ is available, we compare models with different change point locations at the subject level. Subsequently, $\widehat{\boldsymbol{\Psi}}_{ij}$ is determined using model selection procedures \citep[see e.g.,][]{Chen1998}.
At the subject level, the same data are used for determining the change point locations as well as estimating the HRs in each stationary segment. This introduces a model selection bias, which needs to be account for at the group level. To this end, we define the post selection variance. 
In Section \ref{subsec:posi}, we present general results regarding the post selection variance for parameters in a regression model with autocorrelated error terms. These results are used in Section \ref{subsec:subject_posi} and \ref{subsec:group_posi}, in which we propose a procedure to test for changes in the HRs when the change point locations are unspecified.

\subsection{Post selection variance}
\label{subsec:posi}
To define the notion of post selection variance, we make use of post selection confidence distributions. Therefore, we first give a short overview of post selection confidence distributions as introduced in \citet{Garcia2023} for independent data. Then, we present results for the construction of post selection confidence distributions when the data under consideration follow an AR(1) model. The proofs are provided in \ref{sec:proof}. We end this subsection with the definition of the post selection variance.

A confidence distribution is a summary of information learned from the data about the parameter of interest under the assumed model \citep{SchwederHjort2016}. 
The spread of a confidence distribution relates to the confidence we have in the location of the parameter of interest. A post selection confidence distribution is a confidence distribution conditioned on the model selection event \citep{Garcia2023}. 

Let $\Y\in\mathbb{R}^{T}$ be a vector of random variables following the multivariate normal distribution $\mathcal{N}_T(\boldsymbol{\mu},\sigma^2 \TV)$, where $\TV$ is specified by an $AR(1)$ process with parameter $0<\rho<1$. The mean vector $\boldsymbol{\mu}\in\mathbb{R}^T$ is modeled using gaussian linear models. We do not assume that $\boldsymbol{\mu}$ has indeed a linear structure. Let $\M=\{M_1,\ldots,M_L\}$ be the set of considered gaussian linear models, where model $M_{\ell}$ specifies $\mathbb{E}(\Y|\mathbf{X}_{\ell})=\X_{\ell}\bbeta_{\ell}$,
with $\mathbf{X}_\ell\in\mathbb{R}^{T\times p_\ell}$ and $\bbeta_\ell\in\mathbb{R}^{p_\ell}$, $\ell=1,\ldots,L$.
The working density of $\Y$ under model $M_\ell$ is given in the following lemma.
\begin{lemma}
\label{lemma:Dist_Y}
    Let \begin{equation*}
\G=\begin{pmatrix}
    0&1&0&\ldots&0\\
    0&0&1&\ldots&0\\
    \ldots\\
    0&0&0&\ldots&1
\end{pmatrix}, \Tilde{\G}=\begin{pmatrix}
    1&0&\ldots&0&0\\
    0&1&\ldots&0&0\\
    \ldots\\
    0&0&\ldots&1&0
\end{pmatrix}, \Bar{\G}=\begin{pmatrix}
    1&0&\ldots&0&0\\
    0&0&\ldots&0&1
\end{pmatrix},\end{equation*}
with $\G, \Tilde{\G}\in\mathbb{R}^{(T-1)\times T}$ and $\Bar{\G}\in\mathbb{R}^{2\times T}$. Under the gaussian linear model $M_\ell$ the joint density of $\Y$ in its natural parametrization is given by 
\begin{equation}
    f(\Y|\mathbf{X}_\ell,\bbeta_\ell,\sigma, \TV)= 
\exp\{\divsig\bbeta_\ell^\top\X_\ell^\top\V \Y-\divsigtwice \Y^\top\V \Y-\kappa(\X_{\ell},\bbeta_{\ell},\sigma^2,\TV)\},\label{eq:Dist_Y}
\end{equation}
with \begin{equation*}
\kappa(\X_{\ell},\bbeta_{\ell},\sigma^2,\TV)=\divsigtwice(\X_{\ell}\bbeta_{\ell})^\top\V\X_{\ell}\bbeta_{\ell}+\frac{1}{2}\log(|\sigma^2\TV|)+\frac{T}{2}log(2\pi)\nonumber.
\end{equation*}
Since $\Y$ follows an AR(1) process, Eq.~\eqref{eq:Dist_Y} is equivalent to
\begin{equation}
  f(\Y|\mathbf{X}_\ell,\bbeta_\ell,\sigma,\TV) =\exp\{\boldsymbol{\lambda}_{\ell}^\top \boldsymbol{w}_{\ell} - \kappa(\X_{\ell},\bbeta_{\ell},\sigma^2,\TV)\}\label{eq:Dist_Y_1}
\end{equation} with natural parameters \begin{equation*}
    \boldsymbol{\lambda}_{\ell}=(\divsig\bbeta_{\ell}^\top,\divsig\sirho\bbeta_{\ell}^\top, -\divsig\sqrho\bbeta_{\ell}^\top,-\divsigtwice, -\divsig\sirho,\divsigtwice\sqrho)^\top,
\end{equation*} and corresponding sufficient statistics \begin{align*}
    \boldsymbol{w}_{\ell}=(&(\X_{\ell}^\top \Y)^\top,[(\Tilde{\G}\X_{\ell})^\top \G \Y+(\G\X_{\ell})^\top\Tilde{\G}\Y]^\top, [2\X_{\ell}^\top \Y- (\Bar{\G}\X_{\ell})^\top\Bar{\G}\Y]^\top,\\
    & \Y^\top\Y,(\Tilde{\G}\Y)^\top\G \Y, [2\Y^\top \Y-(\Bar{\G}\Y)^\top\Bar{\G}\Y])^\top.
\end{align*} 
\end{lemma}

We select one model in $\M$ based on a selection criterion which partitions the sample space $\mathcal{Y}$ in such a way that $\mathcal{Y}=\bigcup_{\ell=1}^L A_\ell$. The set $A_\ell$ denotes the selection region for model $M_\ell$, with $A_{\ell}\cap A_{\ell'}=\emptyset$ if $\ell\neq\ell'$. 
Model $M_{\js}$ is selected if $\Y_{obs}\in A_{\js}$, where $\Y_{obs}$ denotes a realization of $\Y$. 

Post selection confidence distributions are derived for a one-dimensional focus parameter. For the selected model $M_{\js}$, the focus parameter $\theta_{\js}$ is given by some function of the natural parameters $\boldsymbol{\lambda}_{\js}$. Let $\Theta$ denote the parameter space of $\theta_{\js}$. Because we do not assume that $\boldsymbol{\mu}$ has a linear structure, every model in $M$ might be misspecified. Hence, we aim to derive post selection confidence distributions for the pseudo-true parameter $\theta_{\js}'$ \citep{Garcia2023}.

A function $\mathcal{C}_{T,\js}:\Theta\times A_{\js}\to [0,1]:(\theta_{\js},\Y)\mapsto\mathcal{C}_{T,\js}(\theta_{\js},\Y)$ is considered to be a post selection confidence distribution if it satisfies the following two requirements (cf. Definition 3 in \citet{Garcia2023}):
\begin{enumerate}
    \item The function $\mathcal{C}_{T|\js}(\theta_{\js},\Y_{obs})$ is a cumulative distribution function on $\Theta$, for each $\Y_{obs}\in A_{\js}$.
    \item For the pseudo-true parameter $\theta_{\js}'$, $\mathcal{C}_{T|\js}(\theta_{\js}',\Y)$ is a function of $\Y$ and follows a uniform distribution on the interval $[0,1]$, regardless of the value of $\theta_{\js}'$.
\end{enumerate}  
\citet{Garcia2023} obtain post selection confidence distributions under the following assumptions:
\begin{assumption}
	\label{as_considered_models}
	All models in $\M$ have a selection probability larger than zero.
\end{assumption}
\begin{assumption}
	\label{as_selection_region} 
	For all $\ell\in\{1,\ldots,L\}$, $A_\ell$ only depends on the sufficient statistics for the model parameters of model $M_{\ell}$.
\end{assumption}
The latter assumption is met when the selection criterion is based on the likelihoods of the models in $\M$.

The following two propositions extend the work by \citet{Garcia2023} to data following an AR(1) process. For Proposition \ref{prop:whitened}, we assume that the variance-covariance components $\sigma^2$ and $\TV$ are known before model selection. 
In this case, under model $M_{\js}$, we can simplify the natural parameters $\boldsymbol{\lambda}_{\js}$ given in Lemma \ref{lemma:Dist_Y} such that $\boldsymbol{\lambda}_{\js}=((1/\sigma^2)\bbeta_{\js}^\top,-1/(2\sigma^2))^\top$ with corresponding sufficient statistics $\boldsymbol{w}_{\js}=((\X_{\js}^\top\V \Y)^\top,\Y^\top\V\Y)^\top$.

\begin{prop}
\label{prop:whitened}
Let $M_{\js}\in \M$ be the selected model with covariates $\X_{\js}$. We assume that the models in $\M$ and the selection criterion fulfill Assumptions \ref{as_considered_models} and \ref{as_selection_region}. Furthermore, we assume that $\sigma^2$ and $\TV$ are known a-priori.
Without loss of generality, let $\theta_{\js}=\boldsymbol{\lambda}_{\ell}[1]=\bbeta_{\js}[1]/\sigma^2$ be the one-dimensional focus parameter with parameter space $\Theta$. Denote the sufficient statistic for $\theta_{\js}$ by $w=\boldsymbol{w}_{\js}[1]=\X_{\js}[,1]^\top\V\Y$. Let $\theta_{\js}'$ be the pseudo-true parameter value of $\theta_{\js}$. 
Denote the vector of nuisance parameters by $\boldsymbol{\phi}_{\js}=\boldsymbol{\lambda}_{\js}[-1]$ and let $\TW_{\js}=\boldsymbol{w}_{\js}[-1]$ be the corresponding vector of sufficient statistics. Let $\TW_{\M\setminus M_{\js}}$ be the vector of sufficient statistics for the natural parameters $\boldsymbol{\lambda}_{\ell}$, $\ell=1,\ldots,L, \ell\neq \js$, of all non-selected models in $\M$. Let $\TW_{\M}=(\TW_{\js}, \TW_{\M\setminus M_{\js}})$, where duplicate items and $w$ are removed. Denote the observed values of $w$ and $\TW_{\M}$ by $w_{obs}$ and $\TW_{\M, obs}$, respectively. Let $\mathbb{P}_{\js}$ denote the probability measure under the selected model.
 Then,  \begin{equation*}
    %\label{eq:POSI_cd_general_whitened}
     \C{\js}=\mathbb{P}_{\js}(w>w_{obs}|\TW_{\M}=\TW_{\M,obs}, \Y\in A_{\js})
\end{equation*} is the uniformly most powerful post selection confidence distribution for $\theta_{\js}'$. 
\end{prop}
\begin{remark}
    Proposition \ref{prop:whitened} is valid for AR processes of order 2 and higher as well.
\end{remark}
Because we condition on the sufficient statistic of $-1/(2\sigma^2)$, $\C{\js}$ only contains information about $\bbeta_{\js}[1]$ \citep[cf.][]{Garcia2023}.
	
In practice, reliable information about $\sigma^2$ and $\TV$ is not always available. 
Therefore, in Proposition \ref{prop:ar1}, we focus on the case that $\sigma^2$ and $\TV$ are unknown. 

\begin{prop}
\label{prop:ar1}
Let $M_{\js}\in \M$ be the selected model with covariates $\X_{\js}$. We assume that the models in $\M$ and the selection criterion fulfill Assumptions \ref{as_considered_models} and \ref{as_selection_region}. 
Without loss of generality, let $\theta_{\js}=\boldsymbol{\lambda}_{\ell}[1]=\bbeta_{\js}[1]/\sigma^2$ be the one-dimensional focus parameter with parameter space $\Theta$. Denote the sufficient statistic for $\theta_{\js}$ by $w=\boldsymbol{w}_{\js}[1]=\X_{\js}[,1]^\top\Y$. Let $\theta_{\js}'$ be the pseudo-true parameter value of $\theta_{\js}$. Denote the vector of nuisance parameters by $\boldsymbol{\phi}_{\js}=\boldsymbol{\lambda}_{\ell}[-1]$ and let $\TW_{\js}=\boldsymbol{w}_{\ell}[-1]$ be the corresponding vector of sufficient statistics. Let $\TW_{\M\setminus M_{\js}}$ be the vector of sufficient statistics for the natural parameters $\boldsymbol{\lambda}_{\ell}$, $\ell=1,\ldots,L$, $\ell\neq \js$, of all non-selected models in $\M$. Let $\TW_{\M}=(\TW_{\js}, \TW_{\M\setminus M_{\js}})$, where the sufficient statistics for $\theta_{\js}, \theta_{\js}\rho/(\rho^2-1)$ and $-\theta_{\js}\rho^2/(\rho^2-1)$, as well as duplicate items are removed. Let $\BW_{\js}=(\mathbf{I}_T-\X_{\js}(\X_{\js}^{\top} \X_{\js})^{-1}\X_{\js}^{\top})\Y$. Denote the observed values of $w$, $\TW_{\M}$ and $\BW_{\js}$ by $w_{obs}$, $\TW_{\M, obs}$ and $\BW_{\js,obs}$, respectively. Let $\mathbb{P}_{\js}$ denote the probability measure under the selected model.
Then, \begin{equation*}
%\label{eq:POSI_cb_AR}
    \C{\js}=\mathbb{P}_{\js}(w>w_{obs}|\TW_{\M}=\TW_{\M,obs}, \BW=\BW_{\js,obs}, \Y\in A_{\js})
\end{equation*} is the uniformly optimal post selection confidence distribution for $\theta_{\js}'$.
\end{prop}

The post selection confidence distribution of a focus parameter is used to obtain the respective post selection variance. The post selection variance of focus parameter $\theta_{\js}$ is defined in the following.
\begin{defi}
\label{def:posi_var}
Without loss of generality, let $\theta_{\js}=\bbeta_{\js}[1]/\sigma^2$ be the one dimensional focus parameter in a selected model $M_{\js}$. The post selection variance $\sigma^2_{\theta_{\js}}$ is given by
\begin{equation*}
     \sigma^2_{\theta_{\js}}= \int_{-\infty}^\infty \theta_{\js}^2 \frac{\partial}{\partial \theta_{\js}} \C{\js} d\theta_{\js}- \Big(\int_{-\infty}^\infty \theta_{\js}\frac{\partial}{\partial \theta_{\js}} \C{\js} d\theta_{\js}\Big)^2,
\end{equation*}
where $\C{\js}$ can be deduced from Proposition \ref{prop:whitened} (if $\sigma^2$ and $\TV$ are known) or from Proposition \ref{prop:ar1} (if $\sigma^2$ and $\TV$ are unknown).
\end{defi}
If it is not possible to explicitly obtain the post selection confidence distribution, and hence the post selection variance, we can approximate it. We describe a numerical approximation approach in \ref{sec:Approx_cd}.

Next, we discuss the utilization of the post selection variance in the investigation of changes in the HRs.
\subsection{Subject level analysis}
\label{subsec:subject_posi}
We derive the procedure for undefined change point locations under Assumptions \ref{as:confound_stationary}-\ref{as:number_cp_fixed} in Section \ref{sec:methodology_known_cp}. 
The post selection confidence distribution- and hence the post selection variance- is obtained for a one-dimensional focus parameter. Therefore, we model the HRs in such a way that regression coefficients correspond directly to changes in the shape parameter of interest. Any HRF model that meets this requirement can be used. In this work, we are interested in changes in the peak magnitude and utilize the canonical HRF.
In contrast to the previous sections, we do not assume that the BOLD signal has indeed a linear structure.
The working model of the BOLD signal with changes at the (unknown) time points $\boldsymbol{\Psi}_{ij}=\{\boldsymbol{\psi}_{ij1},\ldots,\boldsymbol{\psi}_{ijK}\}$ is given by \begin{align}
\label{eq:BOLD_nonstat_canonical}
    \boldsymbol{y}_{ij}&=\sum_{k=1}^{K}\sum_{c_k=0}^{C_{k}} \boldsymbol{z}_{ijkc_k}\beta_{ijkc_k}+
   \boldsymbol{\Tilde{Z}}_{ij}\boldsymbol{\delta}_{ij}+\boldsymbol{\epsilon}_{ij},\\
    \boldsymbol{z}_{ijkc_k}&= HRF \otimes \boldsymbol{u}_{ik(\psi_{ijk}[c_k]:T)}. \nonumber
\end{align}
Because the canonical HRF only has one basis function, $\boldsymbol{z}_{ijkc_k}\in\mathbb{R}^{T}$. 

To estimate the change point locations $\boldsymbol{\Psi}_{ij}$, we first choose reasonable combinations of change point locations for each condition. The change point locations are restricted to the respective condition onset time points. Denote by $\mathcal{S}_{ij}$ the set of all considered combinations of change point locations.
Let $\Sij=\{\boldsymbol{s}_{ij1},\ldots,\boldsymbol{s}_{ijK}\}\subset \mathcal{S}_{ij}$ denote an arbitrary subset of possible change point locations, with $\boldsymbol{s}_{ijk}=\{s_{ijk1},\ldots,s_{ijkC_{k}}\}$ denoting the considered change point locations for condition $k$. We denote by $M_{\Sij}$ the model specified by Eq.~\eqref{eq:BOLD_nonstat_canonical} with change point locations $\Sij$. The set of all considered models is denoted by $\mathcal{M}_{\mathcal{S}_{ij}}$. We select model $M_{\HS}$ if the observed BOLD signal $\boldsymbol{y}_{obs,ij}$ lies within the selection region $A_{\HS}$. The selection region is obtained as
\begin{equation*}
\label{eq:selection_region}
     A_{\HS}=\{\boldsymbol{y}_{obs,ij}\in\mathbb{R}^T: \mathcal{L}(\HS; \boldsymbol{y}_{obs,ij})> \mathcal{L}(\Sij;\boldsymbol{y}_{obs,ij}),\quad \text{for all }\Sij\in\mathcal{S}_{ij}\setminus \HS\},
\end{equation*} where $\mathcal{L}(\Sij;\boldsymbol{y}_{obs,ij})$ denotes the likelihood under model $M_{\Sij}$. 

After selecting model $M_{\HS}$, we obtain the post selection variance $\sigma^2_{\beta_{\HS,ijkc_k}}$, defined in Definition \ref{def:posi_var}, for $\beta_{\HS,ijkc_k}$, $c_k>0$. Furthermore, point estimates $\hat{\beta}_{\HS,ijkc_k}$ are computed. These point estimates may correspond to the median post selection confidence estimate, the mean of the post selection confidence distribution \citep{Garcia2023}, or the ordinary least squares estimate. 

\subsection{Group level analysis}
\label{subsec:group_posi}
Let $\hat{\boldsymbol{\gamma}}_{jkc_k}=(\hat{\beta}_{\hat{\mathbf{S}}_{1j},1jkc_k},\ldots, \hat{\beta}_{\hat{\mathbf{S}}_{nj},njkc_k})^\top$ and $\hat{\boldsymbol{\gamma}}_{jkc_k}=\eta_{jkc_k}\boldsymbol{1}_n+\boldsymbol{\xi}_{jkc_k}$ as given in Eq.~\eqref{eq:group_level_stat}.
The covariance matrix of $\boldsymbol{\xi}_{jkc_k}$ is given by $\boldsymbol{\Sigma}_{jkc_k}=\sigma^2_{B_{jkc_k}}\mathbf{I}_n+\boldsymbol{\Sigma}_{W_{jkc_k}}$, with $\boldsymbol{\Sigma}_{W_{jkc_k}}=diag(\sigma^2_{\beta_{\hat{\mathbf{S}}_{1j},1jkc_k}},\ldots,\sigma^2_{\beta_{\hat{\mathbf{S}}_{1j},njkc_k}})$.
The null hypothesis of no change in the peak magnitude, $H_{ jkc_k}: \eta_{jkc_k}=0$, $c_k>0$, is tested for each ROI, each condition, and each change point. As in Section \ref{sec:methodology_known_cp}, the resulting multiple testing problem has a hierarchical structure. Appropriate multiple testing procedures can once again be used to determine a set of null hypotheses to reject.
A summary of the full approach is given in Procedure \ref{proc:unknown_cp}.

\begin{proc}{Undefined change point locations} \\
\label{proc:unknown_cp}
\textbf{Input} 
For each subject $i=1,\ldots,n$, ROI $j=1,\ldots,J$ and condition $k=1,\ldots,K$:
\begin{enumerate}
    \item[-] Pre-processed BOLD signal $\boldsymbol{y}_{obs, ij}$
    \item[-] Condition onset time series $\boldsymbol{u}_{ik}$
    \item[-] Number of change points $C_{k}$
\end{enumerate}
\textbf{Subject level}\\
For each subject $i$ and each ROI $j$:
    \begin{enumerate}
        \item Set $\hat{\boldsymbol{\Psi}}_{ij}=\text{argmax}_{\Sij\subset\mathcal{S}_{ij}}\mathcal{L}(\Sij;\boldsymbol{y}_{obs,ij})$.
        \item For each $k=1,\ldots,K$ and $c_k=1,\ldots,C_k$:
        \begin{enumerate}
            \item[(i)] Compute $\CBOLD{\HS}$. 
            \item[(ii)] Compute $\hat{\sigma}^2_{\beta_{\HS,ijkc_k}}$. 
            \item[(iii)] Compute  $\hat{\beta}_{\HS,ijkc_k}$. 
        \end{enumerate}
    \end{enumerate}
\textbf{Group level}\\
For each ROI $j$, each condition $k$ and each $c_k=1,\ldots,C_k$:\begin{enumerate}
    \item Estimate $\eta_{jkc_k}$ and $\boldsymbol{\Sigma}_{jkc_k}$.
    \item Compute the p-value corresponding to the null hypothesis $H_{jkc_k}: \eta_{jkc_k}=0$.
\end{enumerate}
Determine a set of null hypotheses to be rejected.\\
\textbf{Output}\\
 Set of change points per condition and ROI at which the peak magnitude of the HR changes significantly
\end{proc}

\begin{remark}
	 Once again, the procedure can be applied to the BOLD signal at the voxel level. ROIs can be determined as explained in Section \ref{sec:fmri_intro}. Once again, we require one null hypothesis per voxel. This null hypothesis may be an elementary null hypothesis or one of its ancestors. 
\end{remark}

\section{Simulation studies}
\label{sec:simulation}
We have evaluated the proposed procedures in two simulation studies. We focus on the power of the procedures in different scenarios. We consider a procedure to be powerful when the expected number of discoveries is large, while the number of false discoveries $\mathscr{V}$ is controlled.
When information about the change point locations is available, we do not only  investigate the power of the proposed procedure, but also its robustness against misspecification of the change point locations. All simulation results have been based on $B=1000$ Monte Carlo repetitions.

\subsection{Pre-specified change point locations}
We have simulated an event-related fMRI experiment with $n=30$ hypothetical subjects. The simulated BOLD signal for one ROI has had a length of $T=500$, with a simulated scanning repetition time $TR=2$ seconds. For the simulated experiment, we are only interested in one ROI.
The simulated experimental design has included $K=2$ conditions with $60$ presented stimuli per condition and $itis$ observation time points between the stimuli onsets. We have simulated one rapid change in the HR to each condition. The change point locations have differed between the $n$ hypothetical subjects. %thus the stationary, segment-specific condition onset time series have differed between the hypothetical subjects as well. 
The $120$ trial onset times have been randomly sampled from the observation time points $1,\ldots,T$, under the restriction that $itis\in\{3,4,5\}$. For each hypothetical subject, the change point locations have been randomly drawn from the respective condition onset times, with the restriction that each stationary segment included at least $15$ onsets. 

We have convoluted the stationary, segment-specific condition onset time series with the HRF to compute the rudimentary HR as in Eq.~\eqref{eq:BOLD_nonstat_cp}. The HRF has been modeled using the FLOBS basis set with $G=3$ basis functions as given in FSL \citep{FSL}. 
The regression coefficients corresponding to the HRs before the change point have been set to $\boldsymbol{\beta}_{bc,i}=(3.2,-6.4,3.2)^\top$, $i=1,\ldots,n$, for both conditions. These values have been chosen so that the HRs before the change point had a similar shape to the canonical HR with a peak magnitude of roughly one. 
The HRs after the change points have been set to be $HR_{ac,i}=((\boldsymbol{\beta}_{bc}[1]+e_i)/\boldsymbol{\beta}_{bc}[1]) \cdot HR_{bc,i}$, such that $\boldsymbol{\beta}_{ac,i}[1]=\boldsymbol{\beta}_{bc}[1]+e_i$. The subject specific effect sizes $e_i$ have been sampled from the normal distribution, i.e., $e_i\sim \mathcal{N}(\bar{e},1)$, independently for each $i=1,\ldots,n$. We have set the group effect size $\bar{e}\in\{-1,-0.5,0,0.5,1,1.5,2,2.5\}$. The group effect size $ \bar{e}$ has differed between the conditions. 

The subject specific noise has been chosen to be white noise, such that $\boldsymbol{\epsilon}_i[t]\sim \mathcal{N}(0,\sigma_{\epsilon_i}^2)$ independently for each $t=1,\ldots,T$. We have set $\sigma_{\epsilon_i}^2 = \bar{\boldsymbol{y}}_{clean,i}/SNR$ \citep{neuRosim2011}, with $SNR\in\{1,2\}$. We denote by $\bar{\boldsymbol{y}}_{clean,i}$ the average signal of the simulated BOLD time series before adding the noise.

For the analysis of the simulated data, the change point locations have been assumed to be known. We differentiate between two scenarios. In the first scenario, the given change point locations have been correctly specified. In the second scenario, some change point locations have been misspecified. The deviation from the true change point has been measured in condition onsets and has been randomly sampled from $Unif[-5,5]$, independently for each subject. 
In both scenarios we have applied Procedure \ref{proc:known_cp} to investigate changes in seven shape parameters, namely the peak magnitude ($PM$), nadir amplitude ($NA$), area under the curve ($AUC$),  full width at half maximum ($FWHM$), full width at half nadir ($FWHN$), time to peak ($TTP$) and time peak to nadir ($TPN$), see Figure \ref{fig:shape_HRF}. We have estimated the within subject variances of the shape parameters using numerical approximations as described in \ref{sec:est_subject_variance}.
At the group level, we have employed the restricted maximum likelihood (REML) approach to estimate both the regression coefficients, $\eta_{kc_k}^{[q]}$,$q=1,\ldots,7$, as well as the between subject variance. We have been interested in the null hypotheses $H_{k}^{[q]}: \eta_{k2}^{[q]}=\eta_{k1}^{[q]}$ for both conditions and all considered shape parameters. Only $PM$, $NA$ and $AUC$ have exhibited rapid changes. Therefore, if $\bar{e}\neq 0$, the null hypotheses of no change in $PM$, $NA$ and $AUC$ have been false, whereas the null hypotheses corresponding to no change in $FWHM$, $FWHN$, $TTP$ and $TPN$ have been true in all considered scenarios.
To test the null hypothesis $H_{k}^{[q]}$, we have computed two different test statistics; the $Wald$ test statistic ($T_{Wald}$) and the test statistic based on a variance estimate by \citet{KnappHartung2003} ($T_{KH}$), \begin{equation}
\label{eq:test_statistics}
    T_{KH}^{[q]}= \frac{\hat{\eta}_{k2}^{[q]}-\hat{\eta}_{k1}^{[q]}}{\sqrt{SR\cdot \left(\boldsymbol{1}_n^\top\hat{\boldsymbol{\Sigma}}_{k}\boldsymbol{1}_n\right)^{-1}}},\qquad T_{Wald}^{[q]}=\frac{\hat{\eta}_{k2}^{[q]}-\hat{\eta}_{k1}^{[q]}}{\sqrt{\left(\boldsymbol{1}_n^\top\hat{\boldsymbol{\Sigma}}_{k}\boldsymbol{1}_n\right)^{-1}}},
\end{equation} where $\boldsymbol{1}_n^\top\hat{\boldsymbol{\Sigma}}_{k}\boldsymbol{1}_n$ refers to the estimated variance of $\hat{\eta}_{k2}^{[q]}-\hat{\eta}_{k1}^{[q]}$. 
Furthermore, $SR=(\hat{\boldsymbol{\gamma}}_{k2}-\hat{\boldsymbol{\gamma}}_{k1})^\top \boldsymbol{p}_{k} (\hat{\boldsymbol{\gamma}}_{k2}-\hat{\boldsymbol{\gamma}}_{k1})/(n-1)$ and $\boldsymbol{p}_{k}= \hat{\boldsymbol{\Sigma}}_{k}^{-1}-\hat{\boldsymbol{\Sigma}}_{k}^{-1}\boldsymbol{1}_n\left(\boldsymbol{1}_n^\top\hat{\boldsymbol{\Sigma}}_{k}\boldsymbol{1}_n\right)^{-1}\boldsymbol{1}_n^\top \hat{\boldsymbol{\Sigma}}_{k}^{-1}$.
Both test statistics follow the t-distribution with $n-1$ degrees of freedom under the null. For more information about the test statistics, see \cite{Chen2012}.

To determine a set of null hypotheses to reject., we have applied the procedure by \citet{Bogomolov2020} aiming to control the $sFDR$ at level $\alpha=0.05$.
Note that the hierarchical structure of the null hypotheses had two levels: the two conditions at the first level and the seven shape parameters at the second level.  

The observed number of rejections and the observed FDP in iteration $b$ are denoted by $\mathscr{R}_{obs,b}$ and $FDP_{obs,b}$, respectively. The power of the proposed procedure has been approximated using the rejection rates, $\sum_{b=1}^B\mathscr{R}_{obs,b}/B$. The $FDR$ has been approximated by $\sum_{b=1}^B FDP_{obs,b}/B$. 

\begin{figure}
    \centering
    \includegraphics[width=1\linewidth]{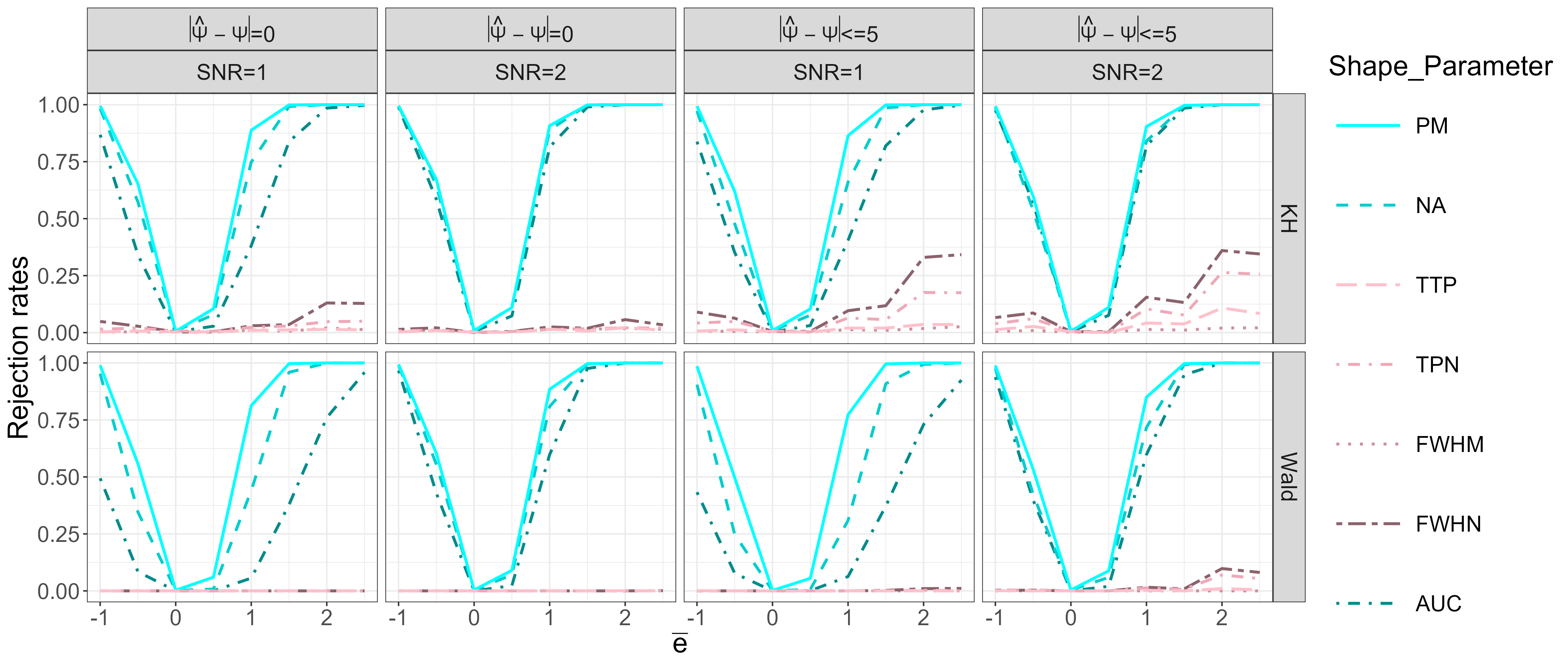}
    \caption{Rejection rates for the null hypotheses of no change for the shape parameters of interest for pre-specified change point locations. The tests have been based on the test statistic inspired by \cite{KnappHartung2003} (upper) and the Wald test statistic (lower) for varying group effect sizes $\bar{e}$ and different signal-to-noise ratios (SNR) under correct specification ($|\hat{\psi}-\psi|=0$) and misspecification ($|\hat{\psi}-\psi|\leq 5$) of the change point location. The rejection rates correspond to the average number of rejected null hypotheses over all repetitions. The null hypotheses of no change in the shape parameters full width at half maximum ($FWHM$), full width at half nadir ($FWHN$), time to peak ($TTP$) and time peak to nadir ($TPN$) (pink lines) are true. For $\bar{e}\neq 0$, the null hypotheses corresponding to no changes in the peak magnitude ($PM$), nadir amplitude ($NA$) and area under the curve ($AUC$) (blue lines) are false. Results are based on $B=1000$ repetitions.}
    \label{fig:RR_known_cp}
\end{figure}

We first present the simulation results regarding the power of the proposed procedure.
Figure \ref{fig:RR_known_cp} displays the rejection rates of the null hypotheses of no change for the different shape parameters under consideration for varying group effect sizes $\bar{e}$ and for different $SNR$. 
The power of the proposed procedure has increased when the absolute value of $\bar{e}$ increased. Furthermore, the power of the procedure has increased when the $SNR$ increased. 
We have observed similar results when using either $T_{KH}$ or $T_{Wald}$, and for correctly and incorrectly specified change point locations. In our simulations, the proposed procedure has been more conservative when using $T_{Wald}$ compared to using $T_{KH}$. When we specified the change point locations incorrectly, the procedure has been slightly less powerful compared to the scenario of correct specification.
We have observed no rejection of a true null hypotheses when the change point locations were correctly specified, $T_{Wald}$ was utilized, and $\bar{e}\neq0$. Using $T_{KH}$, the number of false discoveries, $\sum_{b=1}^B\mathscr{V}_{obs,b}$, has increased when the absolute value of the group effect size, $|\bar{e}|$, increased. For small $SNR$ the number of false discoveries has been greater than for large $SNR$. The number of the false discoveries has been greater when the change point locations were incorrectly specified compared to the case of correct specification. 
 
Next, we investigate $FDR$ control. Table \ref{tab:FDR} displays the average observed $FDP$ for each simulated scenario. Under correct specification of the change point locations, the average observed $FDP$ has been less than $\alpha=0.05$ for almost all scenarios and both test statistics. The only exception has occured when utilizing $T_{KH}$ in the scenario of large $SNR$ and large group effect sizes ($\bar{e}_1=2$, $\bar{e}_2=2.5$). In this case, the average $FDP$ has been only slightly greater than $\alpha=0.05$. 
For the scenario of misspecified change point locations, the average observed $FDP$ has been less than $\alpha$ when $T_{Wald}$ was utilized. This has not been the case for $T_{KH}$, where it has been notably greater than $\alpha$ for large group effect sizes.

These results indicate that $T_{KH}$ should only be utilized when the available information about the change point locations is reliable. Otherwise, one should use $T_{Wald}$ to ensure $FDR$ control.

\begin{table}[ht]
\centering
\begin{tabular}{rrrrrrr}
	\toprule
\hline
&&&\multicolumn{4}{c}{Average $FDP$}\\ \cmidrule{4-7}
 &\multicolumn{2}{c}{Group effect size} &\multicolumn{2}{c}{correct $\Psi$}&\multicolumn{2}{c}{misspecified $\Psi$}\\
SNR & $\bar{e}_1$ & $\bar{e}_2$ &  $KH$ &  $Wald$ & $KH$ & $Wald$ \\ 
  \midrule
  1 & -1 & -0.5 & 0.0211 & 0.0000 & 0.0440 & 0.0002 \\ 
  1 & 0 & 0.5 & 0.0249 & 0.0037 & 0.0312 & 0.0025 \\ 
  1 & 1 & 1.5 & 0.0233 & 0.0000 & \textbf{0.0590} & 0.0003 \\ 
  1 & 2 & 2.5 &\textbf{0.0534} & 0.0000 & \textbf{0.1375} & 0.0035 \\ 
  2 & -1 & -0.5 & 0.0124 & 0.0000 & 0.0446 & 0.0019 \\ 
  2 & 0 & 0.5 & 0.0210 & 0.0047 & 0.0173 & 0.0048 \\ 
  2 & 1 & 1.5 & 0.0177 & 0.0000 & \textbf{0.0751} & 0.0065 \\ 
  2 & 2 & 2.5 & 0.0261 & 0.0003 & \textbf{0.1692} & 0.0406 \\ 
   \bottomrule
\end{tabular}
\caption{Average observed false discovery proportion ($\overline{FDP}=\sum_{b=1}^{B} FDP_b/B$) at the shape parameter level for varying signal to noise ratios ($SNR$) and group effect sizes ($\bar{e}_k$)  under correct specification and misspecification of the change point locations. The test statistic based on \cite{KnappHartung2003} ($KH$) and the $Wald$ test statistic have been utilized to compute p-values. Changes in the HRs to two conditions with different effect sizes were tested at once. We have aimed to control the FDR at level $\alpha=0.05$. Bold values correspond to $\overline{FDP}>\alpha$. Results are based on $B=1000$ repetitions.} 
\label{tab:FDR}
\end{table}

\subsection{Undefined change point locations}
In this simulation study, we investigate the power of the proposed procedure for undefined change point locations for different effect sizes. Numerically approximating the post selection variance is very time-intensive. Therefore, we have generated BOLD signals for $N=500$ hypothetical subjects for each setting. Then, for each of the $B=1000$ repetitions, we have randomly sampled $n=30$ hypothetical subjects without replacement from the $N$ hypothetical subjects for the group level analysis. Therefore, the power of the procedure is approximated by the expected number of discoveries, conditional on the $N$ hypothetical subjects. 

The simulated BOLD signal for each of the $N$ hypothetical subjects has had a length of $T=250$, with $TR=2$ seconds. As above, we have only been interested in one ROI. In the simulated experiment, $60$ stimuli of one condition have been presented. The experimental design has been the same for all $N$ hypothetical subjects. The condition onset times have been randomly sampled from the observation time points in a way that ensured $itis$ observation time point between the onsets, with $itis\in\{3,4,5\}$. We have simulated one rapid change in the HR.

The change point location has been randomly drawn from the condition onset times without the first and last $10$ condition onsets, independently for each hypothetical subject.
The canonical HRF has been convoluted with the condition onset time series to get the HRs per stationary segment, see Eq.~\eqref{eq:BOLD_nonstat_canonical} and below. The regression parameters corresponding to the HR before the rapid change have been set to $\beta_{i0}=1$, $i=1,\ldots,N$. The regression parameters corresponding to the HR after the rapid change have been computed as $\beta_{i1}=1+e_i$, with $e_i\sim \mathcal{N}(\eta, 0.1)$, $\eta\in\{0,0.5,1\}$. A subject-specific baseline $\delta_i$ has been added to the BOLD signal, with $\delta_i\sim \mathcal{N}(10, 1)$. 

The noise at the subject level has been drawn from a multivariate normal distribution, i.e., $\boldsymbol{\epsilon}_i\sim \mathcal{N}_T(\boldsymbol{0}_T,\sigma^2_{\epsilon_i}\mathbf{V})$, where $\mathbf{V}$ models an AR(1) process with parameter $\rho=0.2$.
The variance of the noise has been chosen in such a way that the BOLD signal has had a signal to noise ratio of two.

For the analysis of the simulated data, we have considered four possible change point locations per subject, one of which has been the true change point location. The other considered change point locations have been random draws from the condition onset time series without the first and last ten condition onset times. Additionally, we have ensured that the four considered change points were at least five condition onsets apart.
The subject specific variance $\sigma^2_{\epsilon_i} \mathbf{V}$ has been regarded to be known.
We have used numerical approximation to compute the post selection variance of the focus parameter $\beta_{i1}$ as described in \ref{sec:Approx_cd}. The focus parameter corresponds to the regression coefficient modeling the change in the HR. For a comparison, we have also computed the naive estimate of the within-subject variance $\sigma^2_{\hat{\beta}_{i1}}$, which does not account for model selection.

%\delete{We have considered three different location estimators for the focus parameter, the ordinary least squares (OLS) estimator $\hat{\beta}_{i1}^{[OLS]}$, the median post selection confidence estimate $\hat{\beta}_{i1}^{[0.5]}$ and the expected value of the post selection confidence distribution $\hat{\beta}_{i1}^{[E]}$.}

In total, four different estimation approaches have been compared: the naive approach (naive variance and the ordinary least squares (OLS) estimator $\hat{\beta}_{i1}^{[OLS]}$), the post selection median approach (post selection variance and the median post selection confidence estimate $\hat{\beta}_{i1}^{[0.5]}$, "posi\_05"), the post selection expected value approach (post selection variance and $\hat{\beta}_{i1}^{[E]}$, "posi\_E") and the post selection OLS approach (post selection variance and the expected value of the post selection confidence distribution $\hat{\beta}_{i1}^{[OLS]}$, "posi\_OLS"). 
At the group level, we have used REML to estimate both the regression coefficient, $\eta$ and the between subject variance. The test statistics $T_{KH}$ and $T_{Wald}$, given in Eq.~\eqref{eq:test_statistics}, corresponding to the null hypothesis $H_0: \eta=0$ have been computed. The type I error has been controlled at level $\alpha=0.05$. Since we only consider one ROI, one condition, one change point and one shape parameter, we do not have a multiple testing problem. 

When approximating the post selection variance, in roughly 10\% of cases the procedure failed. We comment on that in \ref{sec:Approx_cd}. The corresponding BOLD time series have not been included in the $N$ simulated BOLD time series.

\begin{figure}
    \centering
    \begin{multicols}{2}
    \includegraphics[width=1\linewidth]{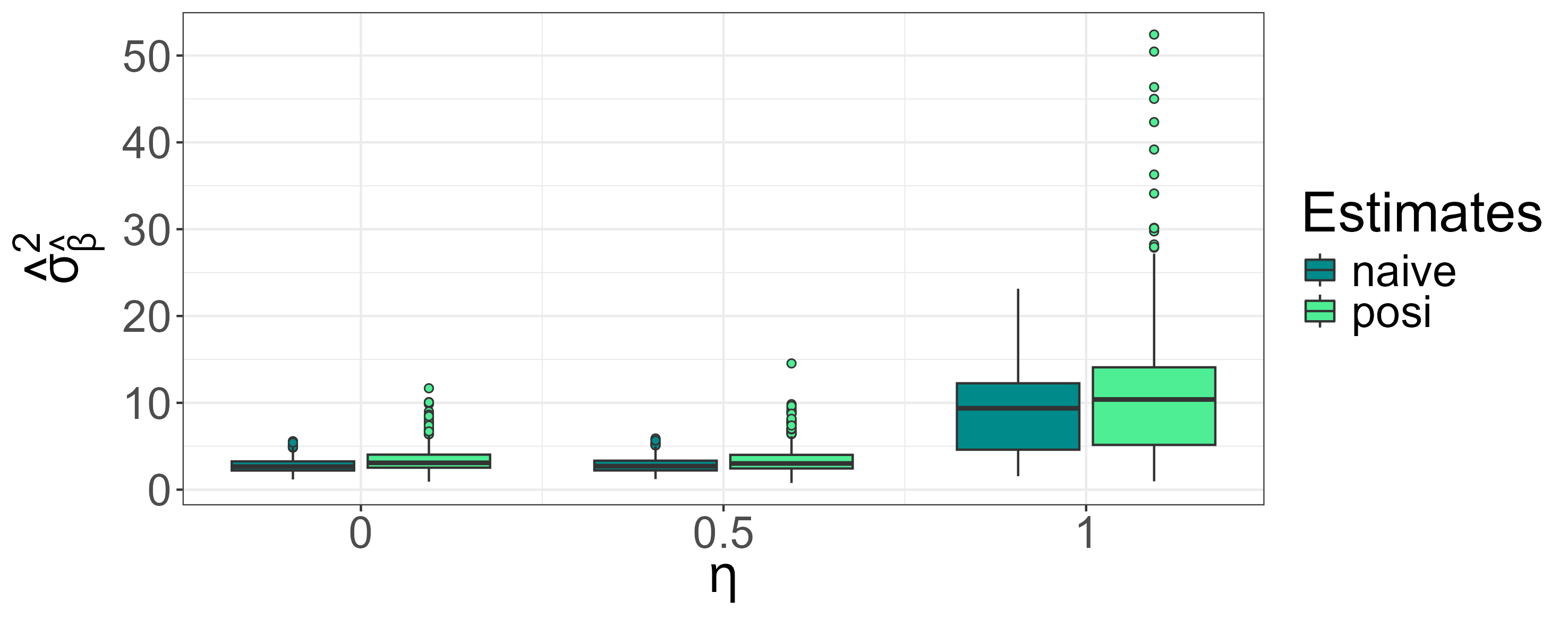}\\
    \includegraphics[width=1\linewidth]{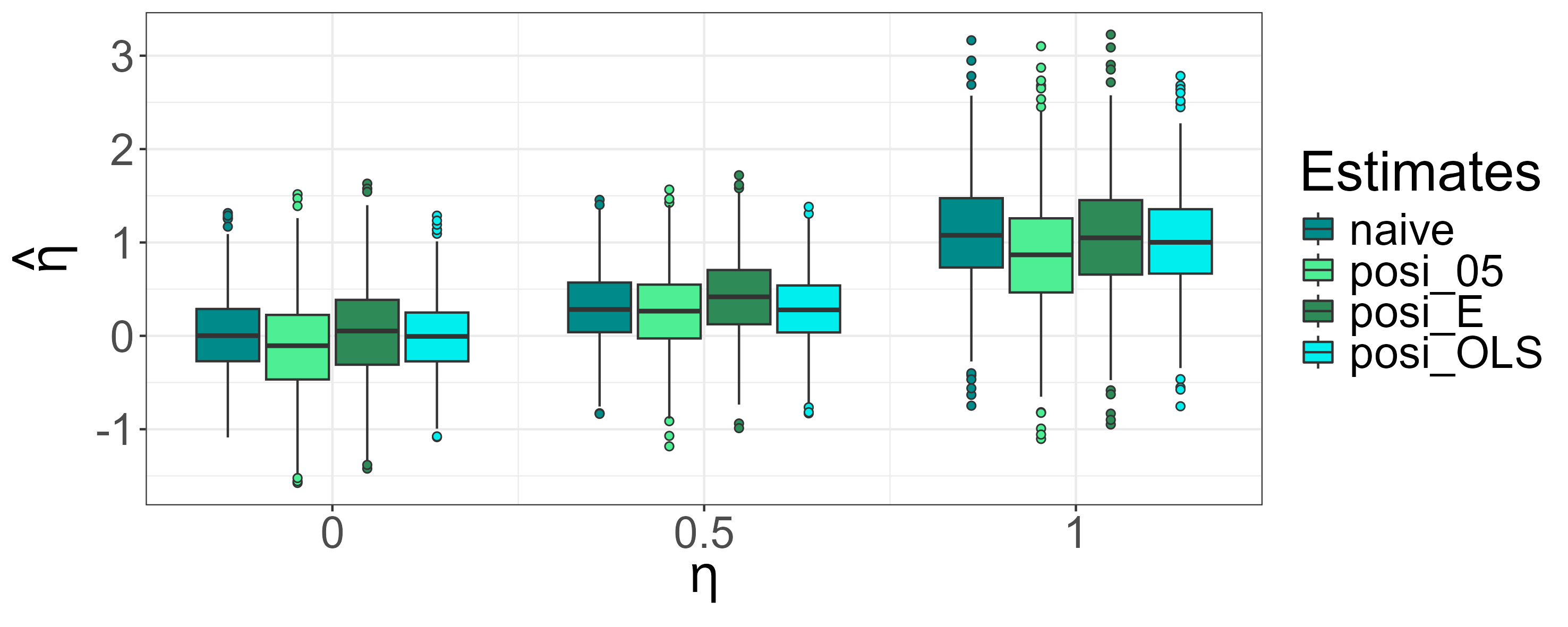}
    \end{multicols}
    \caption{Comparison of estimation approaches for undefined change point locations. The boxplots display the variability and the median of $\hat{\sigma}^2_{\hat{\beta}}$ (left) as well as $\hat{\eta}$ (right) across all iterations for different estimators. Estimators for $\hat{\sigma}^2_{\hat{\beta}}$ are the naive estimator and the post selection (posi) estimator. Regarding $\hat{\eta}$, four different combinations of estimators at the subject level are compared: the naive approach, the post selection mean approach (posi\_05), the post selection expected value (posi\_E) and the post selection OLS (posi\_OLS). Results are based on $B=1000$ repetitions.}
    \label{fig:BP_unknown}
\end{figure}

In the left panel of Figure \ref{fig:BP_unknown}, we compare the simulation results regarding the naive estimate of the within subject variance $\sigma^2_{\hat{\beta}_{i}}$ with the post selection variance. On average, the post selection variance has been larger than the naive variance estimates, with larger variation across the subjects. 
In the right panel of Figure \ref{fig:BP_unknown}, boxplots corresponding to the estimated $\eta$ based on the four different estimation approaches at the subject level are displayed. Overall, the estimation approaches have returned similar results. We have observed that, on average,  $\hat{\eta}$ based on the posi\_E approach has been closest to the true $\eta$. The variance of $\hat{\eta}$ has been largest when the median or expected value of the post selection confidence distribution, i.e., $\hat{\beta}^{[0.5]}$ and $\hat{\beta}^{[E]}$ respectively, were used.

\begin{figure}
    \centering
    \includegraphics[width=1\linewidth]{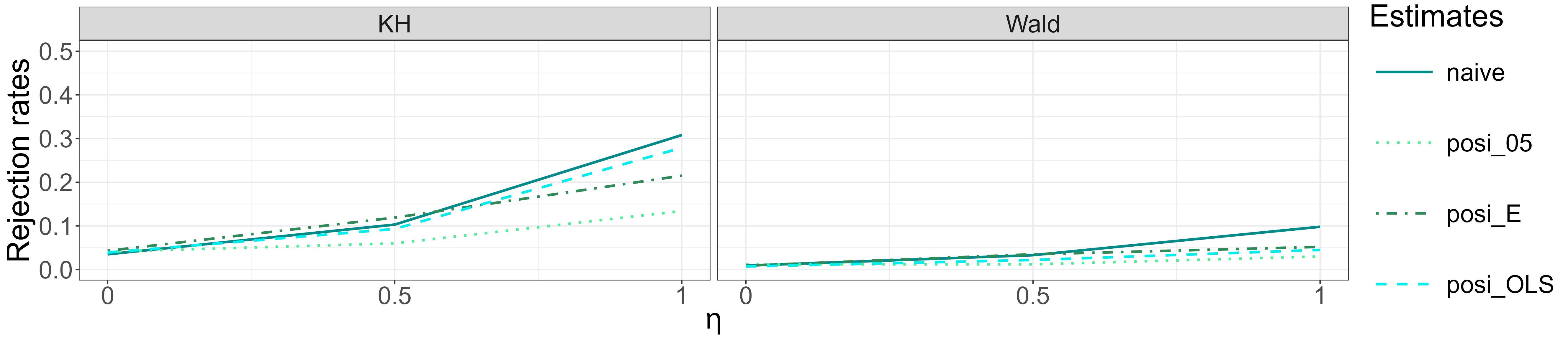}
    \caption{Rejection rates of the null hypothesis of no change in the peak magnitude for undefined change point locations. The tests have been based on the test statistic inspired by \cite{KnappHartung2003} (upper) and the Wald test statistic (lower) for varying group effect sizes $\eta$ . Four different estimation approaches at the subject level are compared: the naive approach (solid line), the post selection mean approach (posi\_05, dotted line), the post selection expected value (posi\_E, dot-dashed line) and the post selection OLS (posi\_OLS, dashed line). The rejection rates correspond to the average number of rejections across all iterations. Results are based on $B=1000$ repetitions.}
    \label{fig:RR_unknown}
\end{figure}

Figure \ref{fig:RR_unknown} displays the rejection rates based on the test statistics $T_{KH}$ and $T_{Wald}$ for the four different estimation approaches at the subject level. In our simulations, utilizing $T_{KH}$ has lead to higher rejection rates than $T_{Wald}$. For all estimation approaches, we have observed rejection rates of less than $\alpha=0.05$ when $H_0$ was true. For smaller group effect sizes ($\eta\leq 0.5$), the posi\_E approach has been the most powerful in our simulations. For larger effect sizes ($\eta=1$), the naive approach has had the highest power.

We emphasize that the simulation results are approximations. Therefore, the observed differences in the performance of the estimation approaches might have occured by chance.

\section{Application: Category learning experiment}
\label{sec:application}
In this section, we make inference on the existence of rapid changes in the shape of the HRs to a category learning task. Data were collected by \citet{WolffBrechmann2022}.
During the experiment, subjects partook in a feedback learning task, aiming to categorize sounds into target vs non-target sounds by pressing one of two buttons, respectively. After each button press, subjects received positive or negative feedback, depending on the correctness of their answer. Each presented sound had five different features, two of which defined the category rule. 
In total, 240 sounds were presented. 
After an initial learning (IL) phase of 120 trials, the buttons' assignment, i.e., target vs. non-target, switched, initiating a relearning (RL) phase. 
The data were acquired in 30 minutes 40 seconds with a repetition time of $TR=2$ seconds and a 3-mm isotropic resolution. The sample size was $n=27$ subjects. 
For more detailed information about the experiment and data acquisition, we refer to \citet{WolffBrechmann2022}.

Pre-processig of the fMRI data has been done using "fmriPrep" (\url{http://github.com/poldracklab/fmriprep}) and has included slice-time and head-motion correction, spatio-temporal filtering, high-pass filtering (discrete cosine filter with 128s cut-off), temporal and anatomical component based noise correction \citep{Behzadi2007} as well as registration to MNI space. Additionally, we have carried out spatial smoothing as described in \citet{Polzehl2019}, Chapter 4.1.9, with a Gaussian filter of 4 $mm$ full width at half maximum.

Fourteen spherical ROIs and the corresponding average BOLD signals have been considered. 
An overview of the ROIs is given in Table \ref{tab:ROI}. The center of the coordinates have been taken from \citet{WolffBrechmann2022} and have been transformed into MNI coordinates using the web-tool provided by  BioImage Suite at \url{https://bioimagesuiteweb.github.io/webapp/index.html}. 
The radius has been fixed at $6$ $mm$, such that each region contains 31 voxels. A visualization of the location of the ROI is given in \ref{sec:ROI}.
\begin{table}[ht]
    \centering
    \begin{tabular}{llrrr}
    	\toprule
         Name& &\multicolumn{3}{c}{Location}  \\
         &&x&y&z\\
         \midrule
         Dorsal Posterior Cingulate Cortex (DPCC) & &0&-35&30\\
         Posterior Medial Prefrontal Cortex (PMPC)& &0&26 &40  \\
         Dorsolateral Prefrontal Cortex (DPC)&L&-44&13&27\\
         &R&44&13&27\\
         Orbitofrontal Cortex (OC)&L&-34&50&15\\                                                              
          &R&34&50&15\\
         Superior Temporal Cortex (STC)&L&-58&-22&-4\\
         &R&58&-22&-4\\
         Anterior Insula (AI) & L & -34&22&-7\\
         & R & 34&22&-7\\
         Caudate Nucleus (CN)&L&-10&5&9\\
          &R&10&5&9\\
         Dorsomedial Thalamus (DT)&L&-8&-11&9\\
         &R&8&-11&9\\
         \bottomrule
    \end{tabular}
    \caption{Description of the spherical regions of interest. Given is the name of the anatomical region in which the region of interest lies and the location of the center in MNI space. The radius for each region is set to $6$ $mm$.}
    \label{tab:ROI}
\end{table}

We have investigated rapid changes in the HRs to the negative and positive feedback conditions. One change point location has been defined between the IL and the RL phase, corresponding to the first trial of the RL phase.
For the HR to positive feedback, we have additionally considered one rapid change during each learning phase, accounting for differences before and after the subjects have potentially discovered the category rule.

This is motivated by \citet{SmithEll2015}, who discuss the issue of modeling rapid changes in the performance of humans during rule-based category learning. Such transitions in learning can be revealed by the backward learning curve \citep{Hayes1953}. To affirm rapid changes in the performance of the subjects in the present experiment, we have constructed such backward learning curves for each half of the experiment. Based on the subject-specific performance, a block of trials has been defined at which a predefined learning criterion is fulfilled for the first time. As learning criterion we have defined twelve consecutive positive feedback trials, which must include at least three correctly specified target tones \citep[c.f.][]{Lommerzheim2023}. Furthermore, to later ensure reliable estimation of the shape parameters, we have required at least nine positive feedback trials before fulfillment of the learning criterion. In other words, the learning criterion is considered to be fulfilled if a positive feedback trial is the first of twelve consecutive positive feedback trials and is additionally preceded by at least nine positive feedback trials, which do not need to be consecutive. The backward learning curves have been constructed by splitting the answering sequence of each subject into blocks of five. The blocks have been numbered consecutively. The block containing the first trial at which the learning criterion is fulfilled has been defined as block zero. After aligning the blocks of all subjects, the average proportion of correct answers for each block has been computed. Subsequently, we have fixed the change point location as the trial at which the above learning criterion is fulfilled for the first time, respectively for each subject and each learning phase.

We have applied Procedure \ref{proc:known_cp} to make inference on the existence and nature of the rapid changes in the HRs to negative and positive feedback. In total, four rapid changes have been investigated, one in the HR to negative feedback and three in the HR to positive feedback. We have modeled the HRF using the FLOBS basis set with the standard parameters defined in FSL \citep{Woolrich2004, FSL}. The serial correlation of the residuals has been assumed to follow an AR(1)-process. Pre-whitening has been carried out as described in \citet{Mumford2006}. 
The shapes of the (estimated) HRs have been described by the seven shape parameters %we have considered in the simulation study, namely 
$PM$, $NA$, $TTP$, $TPN$, $FWHM$, $FWHN$ and $AUC$, see Figure \ref{fig:shape_HRF}. The variances of the shape parameters at the subject level have been approximated as explained in \ref{sec:est_subject_variance}. At the group level, we have used the test statistic $T_{KH}$ as given in Eq.~\eqref{eq:test_statistics} to compute the p-values corresponding to the tested null hypotheses. 

Because we consider several change points in the HR to positive feedback, we cannot assume that the null hypotheses at the condition level are independent. Thus, we have applied the procedure by \citet{GoemanFinos2012} to control the $FWER$ at level $\alpha=0.05$.\\

\begin{figure}
	\centering
	\includegraphics[width=0.75\linewidth]{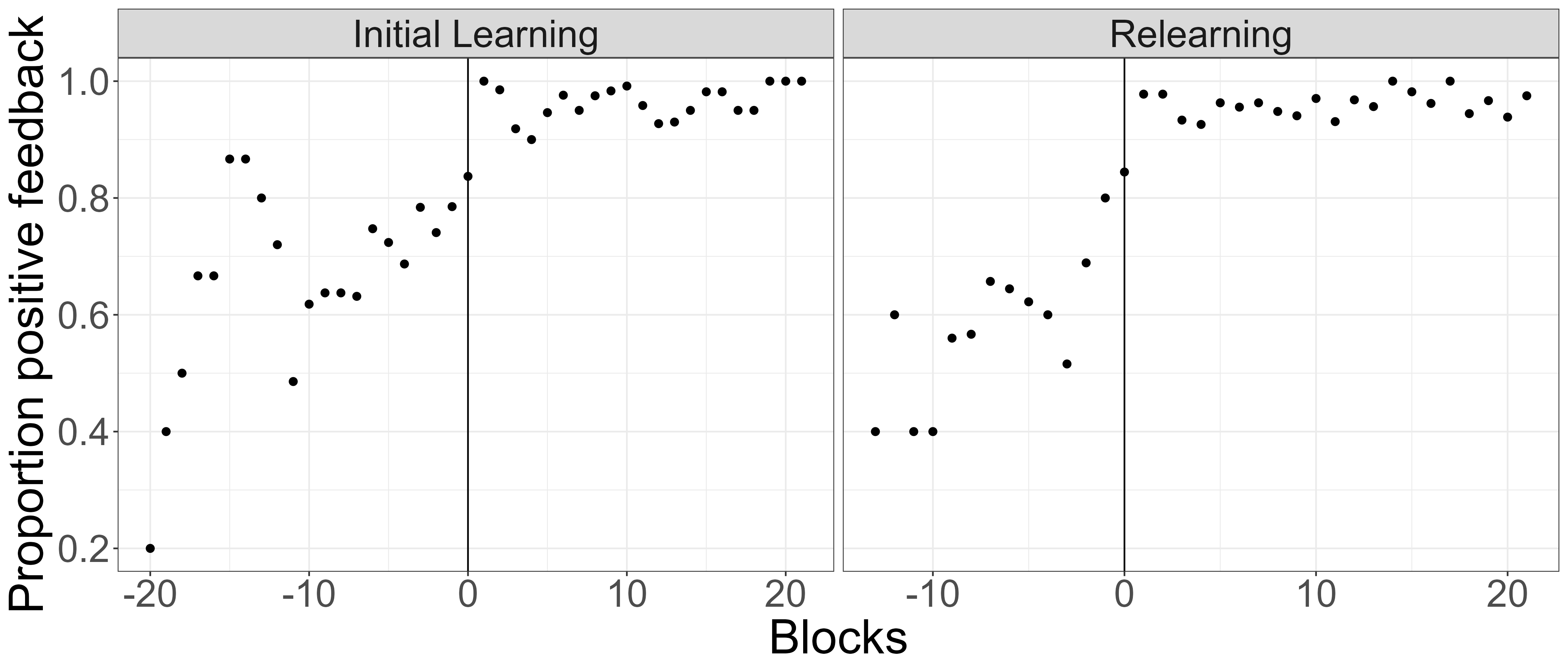}
	\caption{Backward learning curve, i.e., the average proportion of positive feedback trials per block across all $n=27$ subjects. The data per subject are aligned such that the learning criterion is fulfilled during block zero.}
	\label{fig:BLC}
\end{figure}

The backward learning curves of the IL and the RL phase are displayed in Figure \ref{fig:BLC}. They indicate a rapid change in performance and thus support the assumption of rapid changes in the HR due to category learning.

Table \ref{tab:p_vals_roi} displays the p-values corresponding to the null hypotheses of no change in the HR at the ROI level and at the condition level, respectively. Note that the hierarchical structure of the hypotheses is similar to that displayed in Figure \ref{figure:Tree_2}.  
Our procedure identifies significant changes in six different ROI, namely the right DT, the DPCC, the right OC, the left and right AI as well as the PMPC. For the DPCC, the changes could not be attributed to a specific condition. 
That is, though we do find that the HR in the DPCC changes rapidly over time, our method is not able to specify for which condition, change point or shape parameter the HR changes rapidly. 

\begin{figure}
\begin{center}
\includegraphics[width=0.5\linewidth]{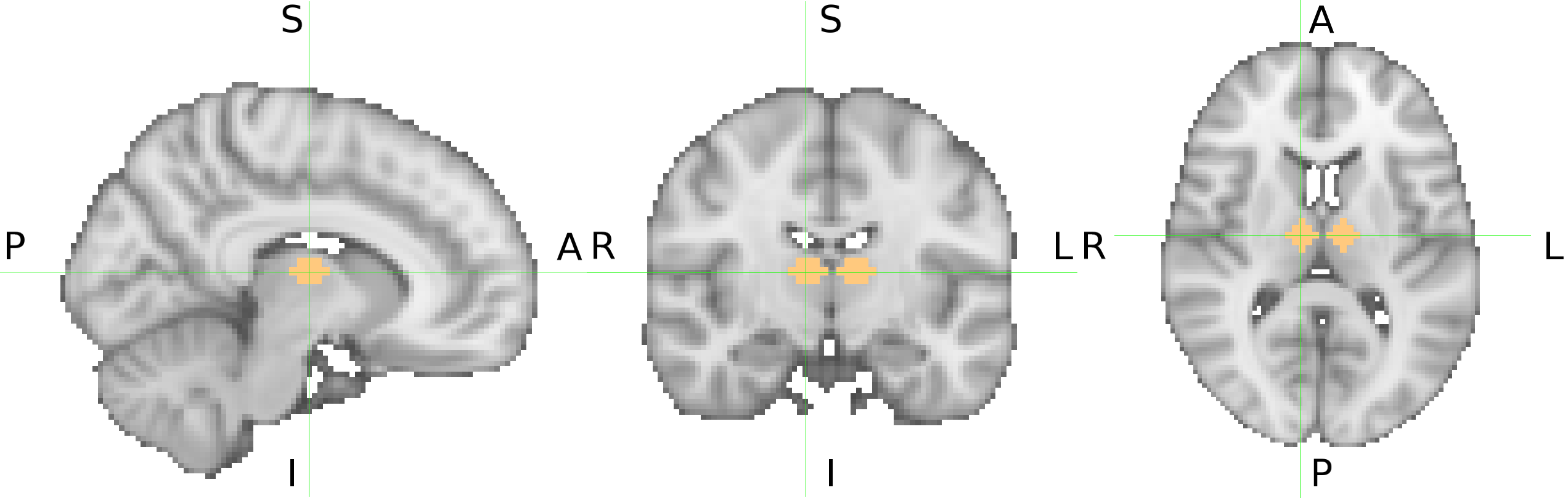}\\
\includegraphics[width=0.65\linewidth]{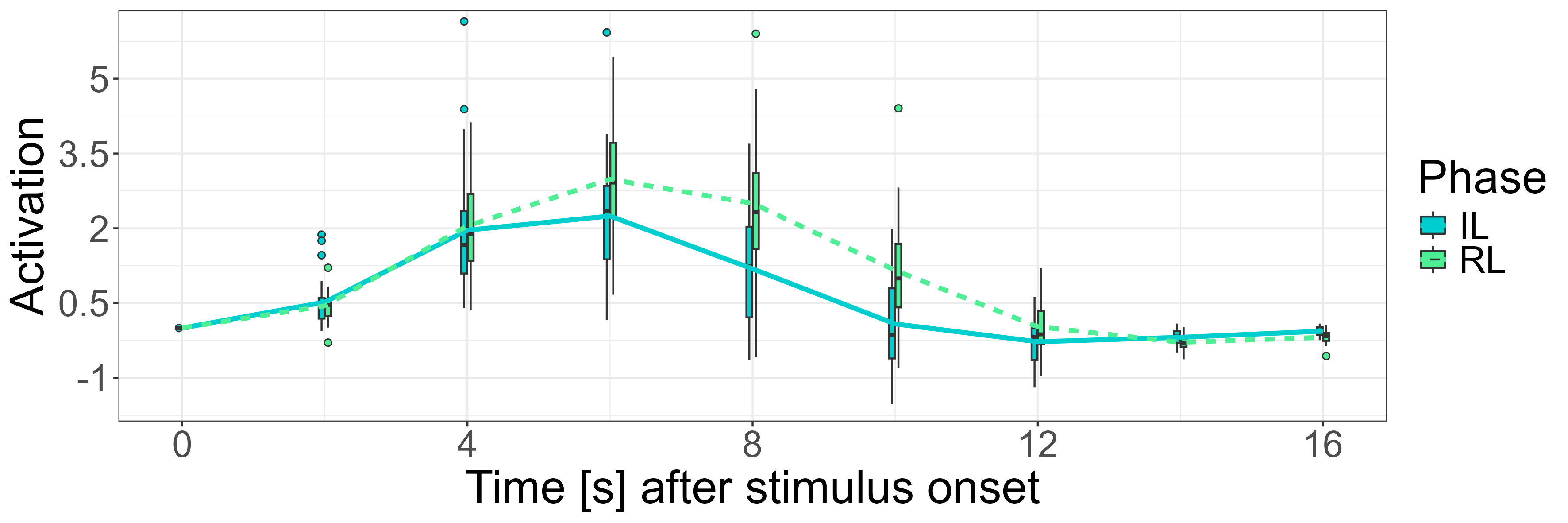}\\
\includegraphics[width=0.5\linewidth]{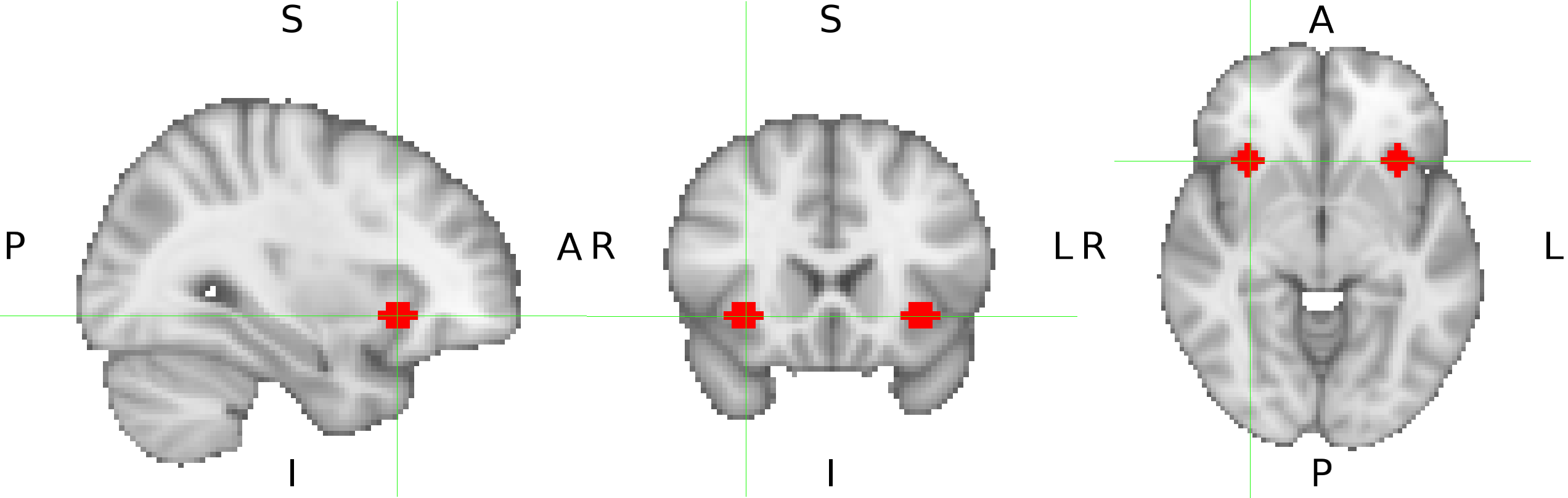}\\
\includegraphics[width=0.8\linewidth]{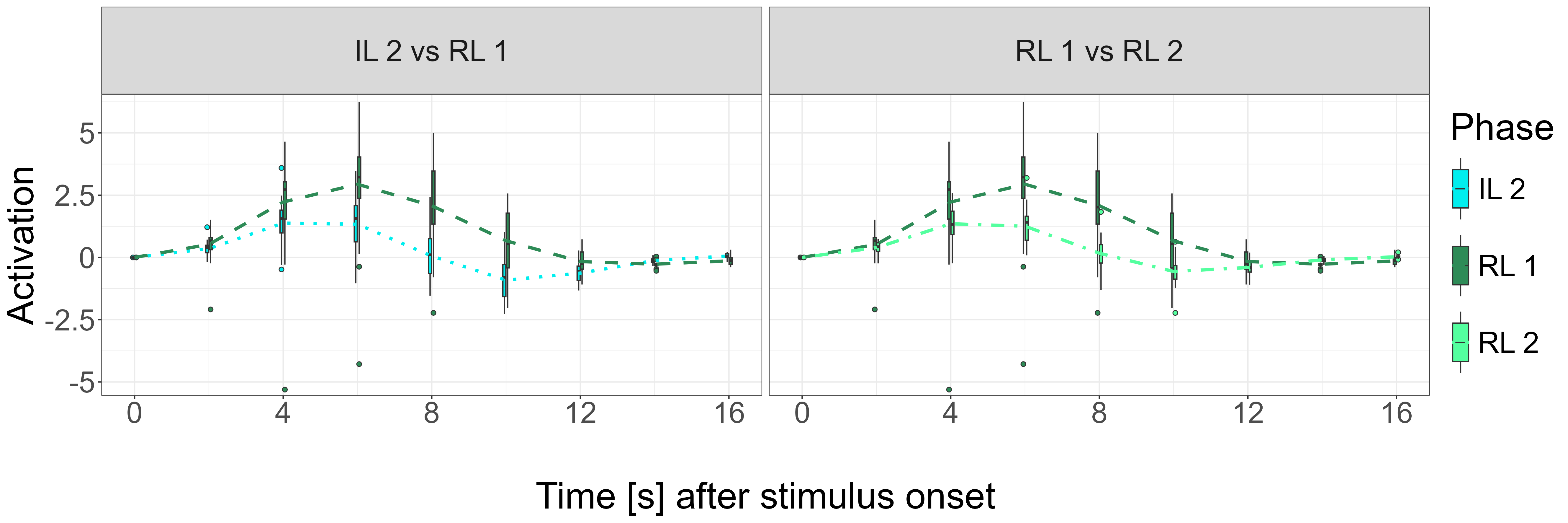}\\
\end{center}
\caption{(Upper two figures) Location of the right and left dorsomedial thalamus (DT) and the average estimated hemodynamic response (HR) across all $n=27$ subjects to the negative feedback condition in the right DT and in the stationary segments.\\ 
(Lower two figures) Location of the right and left anterior insula (AI) and the average estimated HR across all $n=27$ subjects to the positive feedback condition in the right AI and in the stationary segments. Only stationary segments with significant changes in the HRs are compared. \\
The HRs at every $2$ seconds are illustrated. The boxplots display the variation across the subjects.
For negative feedback, the HRs in the initial learning (solid blue line) and relearning (dashed green line) phase are displayed. For positive feedback, the displayed stationary segments correspond to the initial learning phase after rule discovery (IL 2, dotted blue line), relearning phase before rule discovery (RL 1, dashed dark green line) and relearning after rule discovery (RL 2, dot-dashed green line). Only HRs with significant differences are compared.
}
\label{fig:DT_AI}
\end{figure}

Table \ref{tab:results_fmri} displays the p-values corresponding to the rejected elementary null hypotheses. The HR to negative feedback differs in the right DT. Specifically, the shape parameter $TTP$ differed significantly between the IL and RL phase, with longer $TTP$ in the RL phase.
The corresponding estimated HR to negative feedback for the two learning phases is displayed in the upper half of Figure \ref{fig:DT_AI}. 
Regarding the HR to positive feedback, we find significant changes between the IL and RL phase, as well as changes within the RL phase in the the right OC, the left and right AI and the PMPC.
Between the learning phases and in the RL phase, the HR to positive feedback differs in up to four shape parameters, depending on the ROI. Namely, between the learning phases and in the right AI, we find significant changes in the shape parameters $PM$, $TTP$, $TPN$ and $FWHM$. In the RL phase, the HR to positive feedback exhibits rapid changes in the $PM$, $TPN$, $FWHM$ and $FWHN$ in the PMPC. 
The lower two rows in Figure \ref{fig:DT_AI} illustrate the location of the right AI as well as the respective estimated HRs to positive feedback in the stationary segments in which the HR differ significantly.  
A table displaying the p-values corresponding to all elementary null hypotheses whose parent null hypotheses have been rejected is deferred to \ref{sec:ROI}.

\begin{table}
    \centering
    \begin{tabular}{llllr}
    	\toprule
       \multicolumn{3}{c}{Null hypothesis} &  & \\ \cmidrule{1-3}
        ROI & Feedback && p-value&critical value\\
        \midrule
    CN L& -& $H_{ CN\_L}$& 0.02397 & $\alpha/14$\\
    CN R &-& $H_{ CN\_R}$& 02147&$\alpha/14$\\
   
    DPC L &-& $H_{ DPC\_L}$& 0.00741&$\alpha/14$\\
    DPC R  &-& $H_{ DPC\_R}$& 0.00412&$\alpha/14$\\
    DT L &-& $H_{ DT\_L}$& 0.02689&$\alpha/14$\\
    DT R&-& $H_{ DT\_R}$& \textbf{0.00059}&$\alpha/14$\\
    & negative &$H_{ DT\_R,neg}$&\textbf{0.00029}&$\alpha/(14\cdot4)$\\
    &positive &$H_{ DT\_R,pos}$&0.02188&$3\alpha/(14\cdot4)$\\
    DPCC &-& $H_{ DPCC}$& \textbf{0.00254}&$\alpha/14$\\
    &negative &$H_{ DPCC,neg}$&0.00127 &$\alpha/(14\cdot4)$\\
    &positive&$H_{ DPCC,pos}$&0.00294&$3\alpha/(14\cdot4)$\\
    OC L&-& $H_{ OC\_L}$& 01416&$\alpha/14$\\
   
    OC R&-& $H_{ OC\_R}$& \textbf{0.00061}&$\alpha/14$\\
    &negative&$H_{ OC\_R,neg}$&0.00421&$\alpha/(14\cdot4)$\\
    &positive&$H_{ OC\_R,pos}$&\textbf{0.0.00031}&$3\alpha/(14\cdot4)$\\
    AI L  &-& $H_{ AI\_L}$& \textbf{0.00012}&$\alpha/14$\\
    &negative& $H_{ AI\_L,neg}$&0.00271&$\alpha/(14\cdot4)$\\
    &positive& $H_{ AI\_L,pos}$&\textbf{0.00006}&$3\alpha/(14\cdot4)$\\
    AI R &-& $H_{ AI\_R}$& \textbf{0.00001}&$\alpha/14$\\
    &negative& $H_{ AI\_R,neg}$&0.00166&$\alpha/(14\cdot4)$\\
    &positive& $H_{ AI\_R,pos}$&\textbf{0.00000}&$3\alpha/(14\cdot4)$\\
    PMPC &-& $H_{ PMPC}$& \textbf{0.00008}&$\alpha/14$\\
    &negative&$H_{ PMPC,neg}$&0.18534&$\alpha/(14\cdot4)$\\
    &positive&$H_{ PMPC,pos}$&\textbf{0.00004}&$3\alpha/(14\cdot4)$\\
    STG L &-& $H_{ STG\_L}$& 0.01685&$\alpha/14$\\
    STG R &-& $H_{ STG\_R}$& 0.02307&$\alpha/14$\\
    \bottomrule
    \end{tabular}
    \caption{The p-values corresponding to the null hypotheses of no change in the hemodynamic response at the region level and at the condition level. The location of the regions of interest (ROI) are specified in Table \ref{tab:ROI}. 
    Critical values have been computed based on the inheritance procedure \citep{GoemanFinos2012}, with $\alpha=0.05$.
   The p-values corresponding to the displayed hypotheses are computed based on the p-values of their respective children hypotheses, following \cite{dickhaus_2014}, Chapter 11.2.2. Results are based on the data from $n=27$ subjects. }
    \label{tab:p_vals_roi}
\end{table}

\begin{table}
    \centering
    \begin{tabular}{llllllll}
    	\toprule
        ROI &Feedback &Change Point & \multicolumn{5}{c}{Shape Parameter} \\
        \hline
      DT R& negative &IL- RL & &$TTP$&&& \\
 
      OC R& positive & IL-RL & $PM$ &&$TPN$&$FWHM$&\\
      && RL & $PM$&&&&$FWHN$ \\
     AI L && IL-RL && $TTP$&&& \\
      && RL & $PM$ &&&&\\
      AI R& positive& IL- RL & $PM$&$TTP$&$TPN$&$FWHM$& \\
      && RL & $PM$&&&& \\
      PMPC& positive & IL- RL & $PM$&&$TPN$&& \\
      && RL & $PM$&&$TPN$&$FWHM$&$FWHN$\\
      \bottomrule
    \end{tabular}
    \caption{Overview of the rejected elementary null hypotheses ($p\leq 0.05/(56\cdot4)$). The location of the regions of interest (ROI) are specified in Table \ref{tab:ROI}. The change point "IL- RL" indicates the differences between the initial learning and relearning phase and "RL" indicates the change within the relearning phase. Results are based on the data from $n=27$ subjects. ($PM$=peak magnitude,$TTP$=time to peak, $TPN$= time peak to nadir, $FWHM$ full width at half maximum, $FWHN$= full width at half nadir)}
    \label{tab:results_fmri}
\end{table}

The results indicate that, at least in the RL phase, the information carried in the positive feedback is processed differently in some ROI depending on whether the category rule has been learned or not. Whereas information during the IL phase is processed similarly before and after learning, the switch in the category rule initiating the RL phase appears to influence the processing of positive feedback as well. 

\section{Discussion}

We have presented two novel procedures to make inference on the existence of rapid changes in the shapes of the HRs at the group level for event-related fMRI experiments. 
Rapid changes are modeled by splitting the corresponding condition onset time series at certain change point locations. In contrast to existing methods, we allow the change point locations to be condition specific, ROI specific and subject specific. 
The first procedure can be applied when reliable information about the change point locations is available. In this case, the HRs are estimated separately for each stationary segment and condition. This allows for a comparison of multiple shape parameters of the HRs at each change point across all subjects. 
The second procedure does not require any knowledge about the change point locations. We propose using model selection approaches to determine the change point locations. To account for the model selection bias, we utilize the post selection variance. 
At the group level, inference is made on the changes in one shape parameter of the HRs. This shape parameter can be chosen by the practitioner under the prerequisite that changes in the shape parameter are modeled by regression coefficients in the GLM of the BOLD signal. 

Through simulations, we have demonstrated that the proposed procedure for pre-specified change point locations is powerful in detecting rapid changes for medium to large group level effect sizes. 
For large $SNR$ and large group level effect sizes, the proposed procedure utilizing $T_{KH}$ might not control the $FDR$. Since we approximate the $FDR$ through simulations, this might be due to the approximation error.
When the change point locations are misspecified, the power of our procedure is reduced. In this case, the $FDR$ is controlled when utilizing $T_{Wald}$. This emphasizes the need for reliable information about the change point locations. The power of the procedure for undefined change point locations has been overall low in our simulations.

We have applied the proposed procedure for pre-specified change point locations to fMRI data of a category learning experiment, collected by \citet{WolffBrechmann2022}. In addition to change point locations due to the experimental design, we have included subject-specific change points corresponding to rule learning. Notably, we have found evidence for rapid changes in the HRs in the ROI DPCC; the same ROI which has been identified by \citet{WolffBrechmann2022} to exhibit rapid changes in the HR to negative feedback. While our methodology has not been able to identify which HR exhibited the rapid change in the DPCC, it has found evidence for rapid changes in several shape parameters of the HRs to positive feedback in multiple other ROIs. This expands the results by \citet{WolffBrechmann2022}. 
Comparability of the results presented in this work and the work by \citet{WolffBrechmann2022} is somewhat limited due to different analysis and modeling approaches. 
Overall, the case study demonstrates the usefulness of the proposed procedure in providing new insights into the behavior of the HR during category learning.

The proposed procedures can be applied whenever it is of interest to make inference on the existence of rapid changes in the shape of the HRs at the group level. 
Whereas the procedure for undefined change point locations requires less prior knowledge, the procedure for pre-specified change point locations enables a more in-depth analysis of the rapid changes. 
Since the procedure for pre-specified change point locations is fast to compute and has high power, it can easily be applied at the voxel level. Therefore, it can be used to determine ROIs in which the HRs display rapid changes. In contrast, the computational complexity, that is, the number of operations and thus the computation time, of the procedure for undefined change point locations is very large. This is due to the approximation of the post selection variance, which requires solving multiple optimization problems. Therefore, it is preferable to apply Procedure \ref{proc:unknown_cp} at the ROI level instead of at the voxel level. Practitioners can reduce the computational costs of the post selection variance. For instance, the set of possible change point combinations should be minimized before analyzing the data.
The computational complexity is further reduced when the variance at subject level is known a-priori. Thus, when designing an experiment with unknown change point locations, practitioners could add a time segment that can be used to estimate the subject-specific variance.

Although our procedures are tailored for event-related fMRI data, the proposed concept of post selection variance is applicable to other fields as well. Indeed, the results presented in Section \ref{subsec:posi} can be used whenever post selection inference for regression models is of interest, given that the underlying data follow an AR(1) process. In addition to neuroscience, fields of application include economics, finance, or climate science.

Reducing the computation time for the post selection variance is subject of subsequent research.
Additionally, future work will address the comparison of multiple shape parameters when the change point locations are unknown. This requires adapting the computation of the post selection confidence distributions. Lastly, a natural question concerns the functional connectivity of the ROIs in which the HRs change rapidly. Therefore, combining the modeling of rapid changes in the HR with functional connectivity analysis, similar to \citet{Park2020}, appears desirable. 

The R code for the simulation studies is available at \url{https://github.com/fpreusse/RapidChangesHR}. The graphics were generated using ggplot \citep[cf.][]{ggplot} and FSL \citep{FSL}. 

\section{Funding sources}
Friederike Preusse gratefully acknowledges funding by the Deutsche Forschungsgemeinschaft (DFG, German Research Foundation)- project number 281474342.
\appendix
\section{Proofs}
\label{sec:proof}
\subsection*{Proof of Lemma \ref{lemma:Dist_Y}}
\label{sec:proof_lemma}
Eq.~\eqref{eq:Dist_Y} refers to the natural parameterization of the density of a multivariate normal distribution.
Since $\TV$ models an AR(1) process, $\TV[s,t]=\rho^{|s-t|}$, $0<\rho< 1$ and
\begin{equation*}
        \V[s,t]=\begin{cases}
            \arho \qquad &\text{ if } s=t, t\in\{1,n\}\\
            \crho \qquad &\text{ if } s=t, t\in\{2,\ldots,T-1\}\\
            \brho \qquad &\text{ if } |s-t|=1\\
            0 &\text{else}.
        \end{cases}
\end{equation*}
    The expression of $f(\Y|\mathbf{X},\bbeta,\sigma, \TV)$ given in Eq.~\eqref{eq:Dist_Y_1} follows.

\subsection*{Proof of Proposition \ref{prop:whitened}}
\label{sec:proof_prop_whitened}
    Let $\Y^{pw}=\TV^{-1/2}\Y$ and $\X_{\js}^{pw}=\TV^{-1/2}\X_{\js}$. According to model $M_{\js}$, $\Y^{pw}\sim \mathcal{N}_T(\X_{\js}^{pw}\bbeta_{\js}, \sigma^2 \boldsymbol{I}_T)$, where $\boldsymbol{I}_T$ denotes the $(T\times T)$ identity matrix. 
    Thus, under model $M_{\js}$, $\Y^{pw}$ is a vector of stochastically independent random variables. Proposition \ref{prop:whitened} follows directly from Proposition 1 in \citet{Garcia2023}.

\subsection*{Proof of Proposition \ref{prop:ar1}}
\label{sec:proof_prop_ar1}
We follow the proof of Proposition 2 in \citet{Garcia2023}. Let $M$ be a given gaussian linear model which specifies $\mathbb{E}(\Y| \X)=\X\bbeta$, with $\X\in\mathbb{R}^{T\times p}$, $\bbeta\in\mathbb{R}^p$. As seen in Lemma \ref{lemma:Dist_Y}, the working density of $\Y$ under model $M$ represents an exponential family density with natural parameters \begin{equation*}
        \boldsymbol{\lambda}=\left(\divsig\sirho\bbeta^\top, -\divsig\sqrho\bbeta^\top, \divsig\bbeta^\top, -\divsig\sirho,\divsigtwice\sqrho,-\divsigtwice\right)^\top
    \end{equation*} 
Without loss of generality, let $\theta=\bbeta[1]/\sigma^2$ be the one dimensional focus parameter and denote by $\boldsymbol{\phi}$ the vector of nuisance parameters, such that \begin{equation*}
    \boldsymbol{\phi}=\left(\divsig\sirho\bbeta[-1]^\top, -\divsig\sqrho\bbeta[-1]^\top, \divsig\bbeta[-1]^\top, -\divsig\sirho,\divsigtwice\sqrho,\divsigtwice\right)^\top.
\end{equation*}
 Since $\sigma^2\TV$ is unknown and needs to be estimated, we additionally condition the working density of $\Y$ on $\BW=\BW_{obs}$ \citep{Garcia2023}. 

Under model $M$, \begin{equation*}
         \divsig \Y^\top(\boldsymbol{I}_T-\Xlong{})\boldsymbol{\mu}=\divsig(\Y^\top\X\bbeta- \Y^\top\Xlong{}\X\bbeta)=0.
    \end{equation*}
Therefore, the working density of $\Y$ given in Eq.~\eqref{eq:Dist_Y_1} can be expressed by \begin{align*}
    \label{eq:Y_with_W}
    f(\Y|\X,\bbeta,\sigma, \TV)= \exp\Bigl\{&\divsig\sirho\bbeta^\top[(\Tilde{\G}\X)^\top \G \Y+(\G\X)^\top\Tilde{\G}\Y]\\
    &-\divsig\sqrho\bbeta^\top[2\X^\top \Y- (\Bar{\G}\X)^\top\Bar{\G}\Y]+\divsig\bbeta^\top\X^\top \Y \nonumber \\
    & -\divsig\sirho (\Tilde{\G}\Y)^\top\G \Y \nonumber\\
    & + \divsigtwice\sqrho[2\Y^\top \Y-(\Bar{\G}\Y)^\top\Bar{\G}\Y]-\divsigtwice \Y^\top \Y \nonumber\\
    & + \divsig \Y^\top(\boldsymbol{I}_T-\Xlong{})\boldsymbol{\mu} \nonumber \\
    &-\kappa(\X,\bbeta,\sigma^2,\TV)\Bigr\}. \nonumber
    \end{align*}
    This working density is still part of the exponential family. The vector of natural parameters $\boldsymbol{\lambda}$ given above is extended by the vector $\boldsymbol{\tau}=\boldsymbol{\mu}/\sigma^2$. Thus, we can express the natural parameters by $\boldsymbol{\lambda}=(\theta,\theta\rho/(\rho^2-1), -\theta\rho^2/(\rho^2-1), \boldsymbol{\phi}, \boldsymbol{\tau})^\top$. 
    The sufficient statistics for $\boldsymbol{\tau}$ are given by $\BW=(\boldsymbol{I}_T-\Xlong{})\Y$.\\
    
    Denote by $\mathbf{w}'=(w, \TW^\top, \BW^\top)^\top$ the vector of sufficient statistics for $(\theta,\boldsymbol{\phi}^\top,\boldsymbol{\tau}^\top)^\top$.
    By Lemma 2.7.2 in \citet{Lehmann2006}, the joint distribution of $\mathbf{w}'$ has density \begin{equation*}
        f_{w,\TW,\BW}(\mathbf{w}')=h'(\mathbf{w}')\exp\{\sum_{i=1}^{1+3p+T}\mathbf{w}'[i]- \kappa(\boldsymbol{\lambda})\}
    \end{equation*} 
    with some function $h'(\mathbf{w}')$ and $\kappa(\boldsymbol{\lambda})=\kappa(\X,\bbeta,\sigma^2,\TV)$ as in Lemma \ref{lemma:Dist_Y}. Similarly, denote by $\mathbf{w}''=(\TW^\top, \BW^\top)^\top$ the vector of sufficient statistics for $(\boldsymbol{\phi}^\top,\boldsymbol{\tau}^\top)^\top$. The joint distribution of $\mathbf{w}''$ is in the exponential family with density \begin{equation*}
        f_{\TW, \BW}(\mathbf{w}'')=h''(\mathbf{w}'')\exp\{\sum_{i=1}^{3p+T}\mathbf{w}''[i] - \kappa(\boldsymbol{\lambda})\}
    \end{equation*}
    Hence, $w$ conditioned on $\mathbf{w}''$ has density \begin{equation}
    \label{eq:R_given_U}
        f_{w|\TW, \BW}(w)=\frac{f_{w,\TW,\BW}(\mathbf{w}')}{f_{\TW, \BW}(\mathbf{w}'')}
        = \frac{h'(\mathbf{w}')}{h''(\mathbf{w}'')}\exp\{\theta w\}.
    \end{equation}
    Since $\TW$ and $\BW$ are sufficient statistics for $\boldsymbol{\phi}$ and $\boldsymbol{\tau}$, respectively, the density displayed in Eq.~\eqref{eq:R_given_U} only depends on the focus parameter $\theta$. 
    
    Now, we consider the case that model $M_{\js}\in\M$ has been selected based on the data.
    The form of the working density $f_{w|\TW_{\js},\BW_{\js}}(w)$ does not change when conditioning on $\TW_{\M}$ instead of $\TW_{\js}$.
     After conditioning on the selection event, $w|(\TW _{\M}=\TW_{\M,obs}, \BW_{\js}=\BW_{\js,obs}, \Y\in A_{\js})$ follows a truncated exponential family distribution. The truncation limits are defined by the selection procedure. Because the selection regions can be expressed in terms of the sufficient statistics corresponding to the models in $\M$ (see Assumption \ref{as_selection_region}), the domain $dom(w|(\TW_{\M}=\TW_{\M,obs}, \BW=\BW_{\js,obs}, \Y\in A_{\js}))$ is fixed after conditioning on $\TW_{\M}$. Let $a$ and $b$ denote the truncation limits determined by the selection procedure. Denote by $F_{w|\TW_{\M}, \BW_{\js}}$ the cumulative distribution function of  $w|(\TW_{\M}=\TW_{\M,obs}, \BW_{\js}=\BW_{\js,obs})$.  The distribution of $w$ conditional on $(\TW_{\M}=\TW_{\M,obs}, \BW_{\js}=\BW_{\js,obs}, \Y\in A_{\js})$ has the form \begin{equation*}
        f_{w|\TW_{\M}, \BW_{\js}, \Y\in A_{\js}}(w)=\frac{f_{w|\TW_{\M}, \BW_{\js}}(w)\cdot \mathbb{I}(a\leq w\leq b)}{F_{w|\TW_{\M}, \BW_{\js}}(b)-F_{w|\TW_{\M},\BW_{\js}}(a)}, 
    \end{equation*} where $\mathbb{I}$ denotes the indicator function.\\
    To stress the dependency of $f_{w|\TW_{\M}, \BW_{\js}, \Y\in A_{\js}}(w)$ on the focus parameter $\theta\in\Theta$, we use in the following $f_{w|\TW_{\M}, \BW_{\js}, \Y\in A_{\js}}(w;\theta)$ and $F_{w|\TW_{\M}, \BW_{\js}, \Y\in A_{\js}}(w;\theta)$ instead of $f_{w|\TW_{\M}, \BW_{\js}, \Y\in A_{\js}}(w)$ and $F_{w|\TW_{\M}, \BW_{\js}, \Y\in A_{\js}}(w)$, respectively. 
    
   The likelihood ratio $LR(\theta_1,\theta_2)=f_{w|\TW_{\M}, \BW_{\js}, \Y\in A_{\js}}(w;\theta_2)/f_{w|\TW_{\M}, \BW_{\js}, \Y\in A_{\js}}(w;\theta_1)$, with $\theta_2>\theta_1\in\Theta$, is given by \begin{equation*}
        LR(\theta_1,\theta_2)=\frac{F_{w|\TW_{\M}, \BW_{\js}, \Y\in A_{\js}}(b;\theta_1)-F_{w|\TW_{\M}, \BW_{\js}, \Y\in A_{\js}}(a;\theta_1)}{F_{w|\TW_{\M}, \BW_{\js}, \Y\in A_{\js}}(b;\theta_2)-F_{w|\TW_{\M}, \BW_{\js}, \Y\in A_{\js}}(a;\theta_2)}\exp\{(\theta_2-\theta_1)w\},
    \end{equation*} which is everywhere increasing in $w$. Then, by Theorem 5.10 of \citet{SchwederHjort2016}, the confidence distribution $\mathcal{C}_{T|\js}(\theta,\Tilde{\mathbf{y}})=1-F_{w|\TW_{\M}, \BW_{\js}, \Y\in A_{\js}}(w;\theta)$ is uniformly optimal.

\section{Numerical approximations of the within subject variance}
\subsection{Pre-specified change point locations}
\label{sec:est_subject_variance}
The following procedure describes an numerical approximation approach for the subject-level variance $\sigma^2_{\hat{\gamma}_{ijk}}$. The change point locations are considered to be known. \\

Let $\hat{\gamma}_{ijk}$ be the estimate of any shape parameter describing the $HR$.
    \begin{enumerate}
    \item Compute $\hat{\boldsymbol{\Sigma}}_{\hat{\boldsymbol{\beta}}_{ijk}}$, the estimated variance-covariance matrix of the regression coefficients given in Eq.~\eqref{eq:BOLD_at_time_t}.
    \item Generate the vector $\boldsymbol{\beta}_{ijk}^*\in\mathbb{R}^G$ through random sampling from the multivariate normal distribution $\mathcal{N}_G(\hat{\boldsymbol{\beta}}_{ijk}, \hat{\boldsymbol{\Sigma}}_{\hat{\boldsymbol{\beta}}_{ijk}})$.
    \item Compute $\gamma^*_{ijk}$ based on $hr_{ijk}^*=\boldsymbol{\mathcal{B}}\boldsymbol{\beta}^*_{ijk}$.
    \item Repeat steps. 2.-3. for a large number of iterations.
    \end{enumerate}
The empirical variance of the generated shape parameters is an estimate for $\sigma^2_{\hat{\gamma}_{ijk}}$.\\
In the simulation study, we have estimated $\sigma^2_{\hat{\gamma}^{[q]}_{ijk}}$ with $10.000$ iterations.
\subsection{Undefined change point locations}
\label{sec:Approx_cd}
The numerical approximation approach for the post selection variance is based on the work by \cite{Lindqvist2005} and has been described in detail in \citet{Garcia2023}, Section 5, and \citet{SchwederHjort2016}, Chapter 8. 
The idea is to generate samples under the given constraints. \begin{enumerate}
    \item Compute the observed sufficient statistics $(w_{obs},\TW_{\M,obs}^\top)^\top$ (or $(w_{obs},\TW_{\M,obs}^\top, \BW_{obs}^\top)^\top$ if the variance is unknown).
    \item Choose a set $E^{[\theta]}$ of possible values for the focus parameter.
    \item For every $\theta_{e}\in E^{[\theta]}$:
    \begin{enumerate}
    \item Generate samples $\Y^*_d$, $d=1,\ldots,D$, using the density given in Eq.~\eqref{eq:Dist_Y_1}. Note that $\Y^*_d$ has length $T$. The design matrix used in this step is $\X_{obs, \js}$. While generating the sample, it needs to be ensured that the constraints $\TW_{\M}^*=\TW_{\M, obs}$, $ \Y^*_d\in A_{\js}$ and, if necessary, $\BW^*=\BW_{obs}$ hold. To this end, optimization algorithms can be used to adapt the values of the nuisance parameters accordingly. If $\Y^*_d\in A_{\js}$ the sufficient statistic for the focus parameter $w^*_d$ based on $\Y^*_d$ is computed.
    \item Obtain the post selection confidence distribution at value $\theta_{e}$ by\\ $\hat{\mathcal{C}}_{T|\js}(\theta_{e})=\sum_{d=1}^D I(w^*_d > w_{obs})/D$. 
    \end{enumerate}
    \item Denote by $\theta_{[1]},\theta_{[2]},\ldots$ the ordered  parameter values $\theta_e\in E^{[\theta]}$, such that $\hat{\mathcal{C}}_{T|\js}(\theta_{[1]})\leq \hat{\mathcal{C}}_{T|\js}(\theta_{[2]})\leq\ldots$. The post selection variance for the focus parameter is approximated by \begin{equation*}
    \left[\sum_{e=1}^{|E^{[\theta]}|} \theta_{[e]}^2\cdot (\hat{\mathcal{C}}_{T|\js}(\theta_{[e]})-\hat{\mathcal{C}}_{T|\js}(\theta_{[e-1]}))\right]-\left[ \sum_{e=1}^{|E^{[\theta]}|} \theta_{[e]}\cdot  (\hat{\mathcal{C}}_{T|\js}(\theta_{[e]})-\hat{\mathcal{C}}_{T|\js}(\theta_{[e-1]}))\right]^2,
\end{equation*} with $\hat{\mathcal{C}}_{T|\js}(\theta_{[0]})=0$.
\end{enumerate}

For some $\theta_e$, it is infeasible to generate a large number of samples $Y^*_d$ under which model $M_{\js}$ is selected. This happens when the probability of selecting $M_{\js}$ is close to zero. 
 
Therefore, the set of possible values for the focus parameter should be limited to reasonable values. Algorithm \ref{algorithm:possible_values_theta} is a suggestion on how to search for reasonable bounds for $E^{[\theta]}$. The performance of the algorithm depends on the starting value $\hat{\theta}^{[1]}$. We suggest using $\hat{\theta}^{[1]}=\hat{\theta}^{[OLS]}$, which worked well in roughly $90\%$ of the simulations. In the other cases, no reasonable bounds for $E^{\theta}$ could be found and therefore the post selection confidence distribution could not be computed. Investigations regarding an algorithm that is robust against different choices of starting values is left for future research.\\
In the simulation study, we have approximated the post selection variance using $D=500$ iterations.

We end with a short comment on the computational complexity of approximating the post selection confidence distribution and thus the post selection variance. The computational complexity decreases as the effect size $\eta$ increased. This is due to conditioning on the model selection in Step 3 (a) above. As $\eta$ increases, so does the probability that $\Y^*_d\in A_{\js}$. Therefore, generating $D$ samples for which $\Y^*_d\in A_{\js}$ requires less iterations when $\eta$ is large compared to when $\eta$ is small. For similar reasons, the computational complexity increases when the SNR decreases. 

\begin{algorithm}
\caption{Algorithm to find reasonable bounds for $E^{\theta}$}
\small{
\SetKwInOut{Input}{Input}
\SetKwInOut{Output}{Output}
\Input{
$\hat{\theta}^{[1]}$: starting value for $\theta_e$;\\
     Nonnegative number $a$: used to update range of $E^{\theta}$;\\
     Nonnegative integer $D^{[E]}$: number of generated samples $\Y^*$;\\
     Stopping criterion $Stop$: stop algorithm if some function $g(\dot{\theta}, \ddot{\theta})=Stop$
}
\hrulefill\\
$\dot{\theta}_0=\hat{\theta}^{[1]}-a$, $\ddot{\theta}_0=\hat{\theta}^{[1]}+a$, $E^{[\theta]}_0=\emptyset, E^{[\theta]}_1=\{\dot{\theta}_0,\ldots, \ddot{\theta}_0\}$\;
$i=1$\;
\While{$g(\dot{\theta}, \ddot{\theta})\neq Stop$}{
$\dot{\theta}_{full}=\min (\theta_e|\theta_e\in  E^{[\theta]}_i)$, $\ddot{\theta}_{full}=\max(\theta_e|\theta_e\in  E^{[\theta]}_i)$\;
\For{$\theta_e \in E^{[\theta]}_i\setminus E^{[\theta]}_{i-1}$}{
$counter = 0$\;
\For{$d=1,\ldots, D^{[E]}$}{
Compute $\Y^*_d$\; 
\lIf{$\Y^*_d\in A_{\js}$}{$counter= counter +1$}
}
\lIf{$counter= 0$}
{$E^{[\theta]}_i=E^{[\theta]}_i\setminus \theta_e$}

}
\lIf{$E^{[\theta]}_i= E^{[\theta]}_{i-1}$}{
break
}
$\dot{\theta}_i=\min (\theta_e|\theta_e\in E^{[\theta]}_i)$, $\ddot{\theta}_i=\max (\theta_e|\theta_e\in E^{[\theta]}_i)$\;
\lIf{$\dot{\theta}_i= \dot{\theta}_{full}$}{
$\dot{\theta}_i=\dot{\theta}_i-a$
}
\lIf{$\ddot{\theta}_i= \ddot{\theta}_{full}$}{
$\ddot{\theta}_i=\ddot{\theta}_i+a$
}
$E^{[\theta]}_{i+1}=\{\dot{\theta}_i,\ldots,\ddot{\theta}_{i}\}$\; $\dot{\theta}=\theta_i$, $\ddot{\theta}=\ddot{\theta_i}$ \;
\lIf{$E^{[\theta]}_{i+1}=E^{[\theta]}_{i}$}{break}
$i=i+1$\;
}
\Return $\dot{\theta}, \ddot{\theta}$}
\label{algorithm:possible_values_theta}
\end{algorithm}
\section{Application: Additional results}
\label{sec:ROI}
\setcounter{table}{0}
\renewcommand{\thetable}{C.\arabic{table}}
\setcounter{figure}{0}
\renewcommand{\thefigure}{C.\arabic{figure}}
The locations of the regions of interest considered in the case study are displayed in Figure \ref{fig:masks}.
\begin{figure}[ht]
    \centering
    \includegraphics[width=0.7\textwidth]{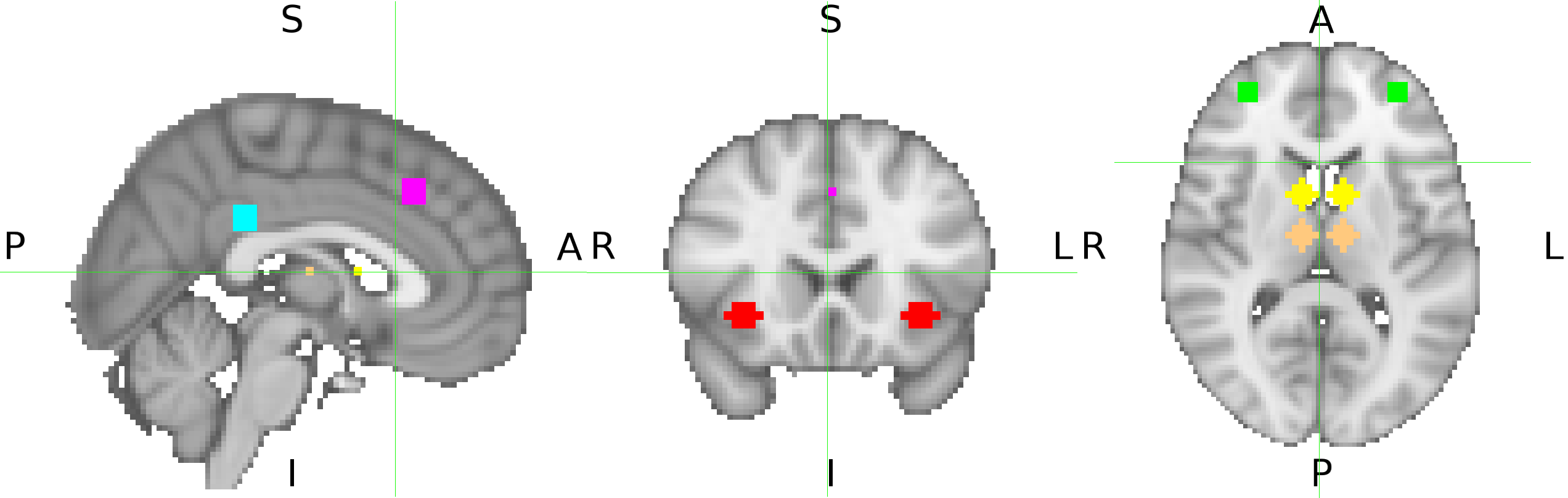}
    \caption{Location of the regions of interestes within the brain. The region in light blue corresponds to the dorsal posterior cingulate cortex, purple to the posterior medial prefrontal cortex, orange to the dorsomedial thalamus, yellow to the caudate nucleus, green to the orbitofrontal cortex and red to the anterior insula. The green lines indicate which cross-sections of the brain are displayed.}
    \label{fig:masks}
\end{figure}

\begin{figure}
	\centering
	\includegraphics[width=0.5\linewidth]{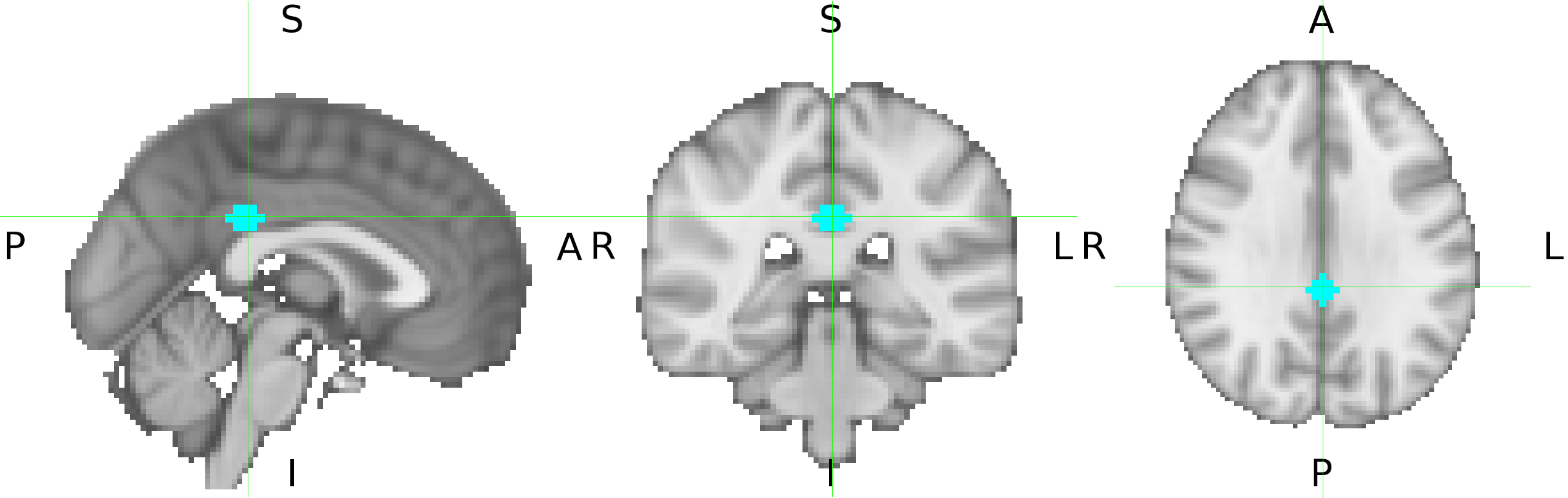}\\
	\includegraphics[width=0.8\linewidth]{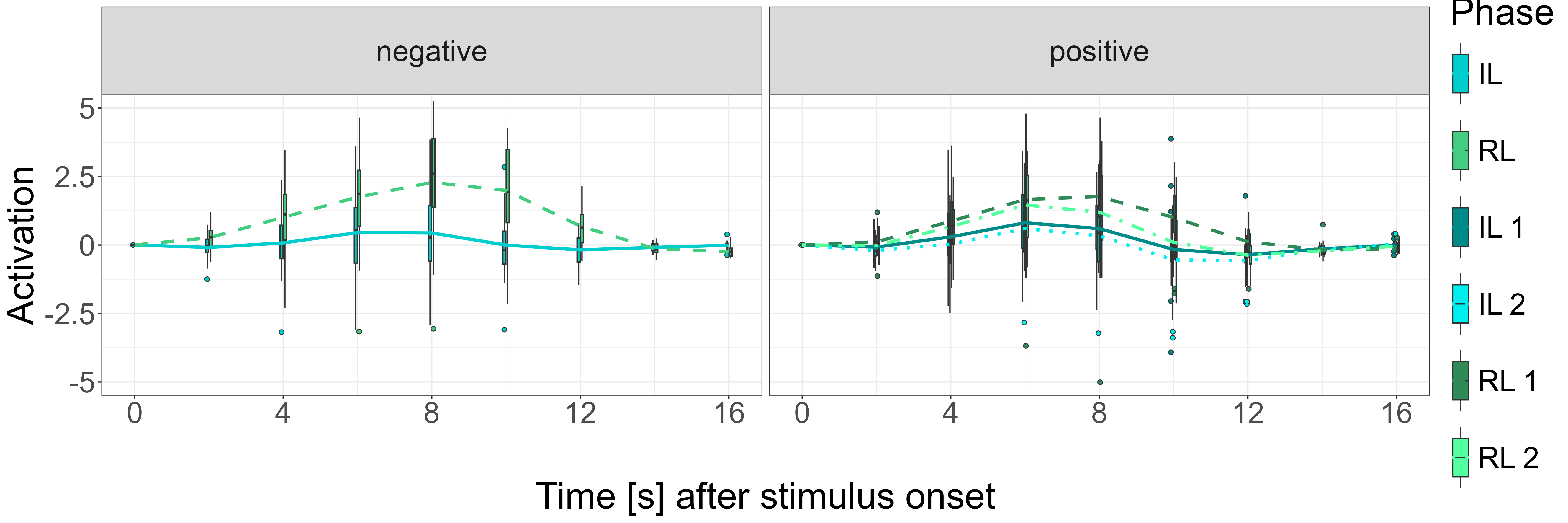}\\
	\caption{Location of the dorsal posterior cingulate cortex (upper) and the average estimated hemodynamic response (HR) across all subjects to negative and positive feedback in the region and in the stationary segments (lower).\\ 
	The HRs at every $TR=2$ seconds are illustrated. The boxplots display the variation across the subjects.\\
	For negative feedback, the HRs in the initial learning (IL, solid blue line) and relearning (RL, dashed green line) phase are displayed. For positive feedback, the stationary segments correspond to the IL before rule discovery (IL 1, solid dark blue line), IL after rule discovery (IL 2, dotted light blue line), RL before rule discovery (RL 1, dashed dark green line) and RL after rule discovery (RL 2, dot-dashed light green line). Results are based on the data from $n=27$ subjects.
	}
	\label{fig:DPCC}
\end{figure}

In our analysis, we have found evidence for changes in the DPCC, but could not allocate this change to the HR to either negative or positive feedback. The HRs to negative and positive feedback in each of the stationary segments in the DPCC are displayed in Figure \ref{fig:DPCC}.

Table \ref{tab:p_val_condition} and \ref{tab:p_vals} display the p-values corresponding to the null hypotheses of no change in the HR at the change point level and at shape parameter level, respectively. Only hypotheses whose parent hypotheses have been rejected are displayed.

\begin{table}
    \centering
    \begin{tabular}{lllll}
    	\toprule
    \multicolumn{4}{c}{Null hypothesis} &  \\ \cmidrule{1-4}
    ROI&Condition& Change point&& p-value  \\ \midrule

    DT R & Neg FB& &$H_{DT\_R,1}$& \textbf{0.00029}\\
    DPCC  & &&$H_{ DPCC,1}$& 0.00127\\
   
    OC R    &&& $H_{OC\_R,1}$& 0.00421\\
    AI L&&& $H_{ AI\_L,1}$& 0.00271\\
    AI R&&&  $H_{ AI\_R,1}$& 0.00166\\
    PMPC& &&$H_{ PMPC,1}$& 0.18534\\

    OC R & Pos FB&IL &$H_{OC\_R,2,1}$& 0.67445\\
     AI L   &&& $H_{ AI\_L,2,1}$& 0.00469\\
    AI R & &  &$H_{ AI\_R,2,1}$& 0.32265\\
    PMPC &  & &$H_{ PMPC,2,1}$& 0.02683\\

    OC R & &IL-RL& $H_{OC\_R,2,2}$& \textbf{0.00010}\\
    AI L&  &  & $H_{ AI\_L,2,2}$& \textbf{0.00070}\\
    AI R & &  &$H_{ AI\_R,2,2}$& \textbf{0.00000}\\
    PMPC &  &  &$H_{ PMPC,2,2}$& \textbf{0.00004}\\

    OC R    & &RL &  $H_{OC\_R,2,3}$& \textbf{0.00022}\\
    AI L &  & & $H_{ AI\_L,2,3}$& \textbf{0.00002}\\
    AI R &  & &$H_{ AI\_R,2,3}$& \textbf{0.00000}\\
    PMPC &  &  &$H_{ PMPC,2,3}$& \textbf{0.00001}\\
    \bottomrule
    \end{tabular}
    \caption{The p-values corresponding to the null hypotheses no change in the hemodynamic response at the change point level. P-values corresponding to rejected null hypotheses are displayed in bold. The critical value at the change point level is $\alpha/(14\cdot4)$, with $\alpha=0.05$. The regions of interes (ROI) are specified in Table \ref{tab:ROI}. The conditions correspond to negative feedback (Neg FB) and positive feedback (Pos FB). The change point "IL- RL" indicates the differences between the learning phases and "RL" indicates the change within the relearning phase. The p-values corresponding to the displayed hypotheses are computed based on the p-values of the respective children hypotheses, following \cite{dickhaus_2014}, Chapter 11.2.2. Results are based on the data from $n=27$ subjects.}
    \label{tab:p_val_condition}
\end{table}

\begin{sidewaystable}
			\begin{tabular}{lllllllllll}
				\toprule
				\multicolumn{10}{c}{Null hypothesis}&    \\  \cmidrule{1-10}
				Cond& CP & ROI & PM&NA&TTP&TPN&FWHM&FWHN&AUC &critical value\\
				\midrule
				Neg FB&IL-RL &DT R&0.03358&0.76915& \textbf{0.00004}&0.03369&0.00107&0.27769&0.99825&$0.05/(56\cdot6)$\\
				Pos FB& IL-RL&OC R&\textbf{0.00007}&&& \textbf{0.00001} &\textbf{0.00005} & & & $0.05/(56\cdot6)$\\
				&&&&0.096294&0.02598&&&0.04501&0.29028& $0.05/(56\cdot4)$\\
				
				&&	AI L& 00143&0.03758&0.00091&\textbf{0.00010}&0.00021&0.17098&0.08238 &$0.05/(56\cdot6)$\\
				
				&&AI R&\textbf{0.00000}&&\textbf{0.00006}&\textbf{0.00009}&\textbf{0.00001}&&&$0.05/(56\cdot6)$\\
				&&&&0.015680&&&&0.15392&0.21809&$0.05/(56\cdot3)$\\
		
			&&	PMPC&\textbf{0.00010}&&&\textbf{0.00001}&&&&$0.05/(56\cdot6)$\\
				&&&&0.14154&0.01261&&0.00472&0.41000&0.36263&$0.05/(56\cdot5)$\\
	
			Pos FB&RL&	OC R& \textbf{0.00005} &&&&&\textbf{0.00003} &&$0.05/(56\cdot6)$\\
				&&	&  &0.32253&0.06627&0.0.00114&0.00110& &0.36152&$0.05/(56\cdot5)$\\
				&&AI L&\textbf{0.00000} &0.52696 &0.07693 &0.00413 &0.03135& 0.13875&0.59612 &$0.05/(56\cdot6)$\\
			
				&&AI R&	\textbf{0.00000} &0.04310 & 0.00175 &  0.00018 & 0.00023 & 0.00075 &0.02631&$0.05/(56\cdot6)$\\
				
			&&	PMPC&	 \textbf{0.00009} &&&\textbf{0.00000} & & \textbf{0.00001} && $0.05/(56\cdot6)$ \\
			&&&&&&&\textbf{0.00022}&&&$0.05/(56\cdot4)$\\
				&&&0.63482&0.06360&&&&&0.0.14404&$0.05/(56\cdot3)$\\
				\bottomrule
			\end{tabular}	
			\caption{The p-values and the critical values corresponding to the elementary hypotheses of the hierarchical structure. The elementary hypotheses are specified by the Condition (Cond), the change point (CP), the region of interest (ROI), and the shape parameters. The conditions are either negative feedback (Neg FB) or positive feedback (Pos FB). The change point "IL- RL" indicates the differences between the learning phases and "RL" indicates the change within the relearning phase. The ROIs are specified in Table \ref{tab:ROI}. 
			Only the children hypotheses of rejected parent hypotheses are displayed.
			P-values corresponding to a rejected null hypothesis are displayed in bold.
			Critical values have been computed based on the inheritance procedure \citep{GoemanFinos2012}. The critical value for a family of elementary hypotheses is adapted once several elementary hypotheses within this family have been rejected. The critical values used for the decision of (not) rejecting the corresponding null hypotheses are displayed. Results are based on the data from $n=27$ subjects. ($PM$=peak magnitude,$TTP$=time to peak, $TPN$= time peak to nadir, $FWHM$ full width at half maximum, $FWHN$= full width at half nadir)
			}
			\label{tab:p_vals}
\end{sidewaystable}

\bibliography{references}
\end{document}